\documentclass[prl,twocolumn,superscriptaddress,hidelinks]{revtex4-2}
\usepackage{listings}
\usepackage{epsfig}
\usepackage{multirow}
\usepackage{amsmath,amssymb,mathtools}
\usepackage{xcolor}
\usepackage{graphicx}
\usepackage{dcolumn}
\usepackage{bm}
\usepackage{subfigure}
\usepackage[utf8]{inputenc}
\usepackage[T1]{fontenc}
\usepackage{tikz}
\usepackage{bigints}
\usepackage{orcidlink}
\usepackage{url}

\usepackage{hyperref}

\begin{document}

\title{Thermoelectric signatures of order-parameter symmetries in iron-based superconducting tunnel junctions}

\author{Claudio Guarcello\,\orcidlink{0000-0002-3683-2509}}
\email{Corresponding author: cguarcello@unisa.it}
\affiliation{Dipartimento di Fisica ``E.R. Caianiello'', Universit\`a di Salerno, Via Giovanni Paolo II, 132, I-84084 Fisciano (SA), Italy}
\affiliation{INFN, Sezione di Napoli Gruppo Collegato di Salerno, Complesso Universitario di Monte S. Angelo, I-80126 Napoli, Italy}
\author{Alessandro Braggio\,\orcidlink{0000-0003-2119-1160}}
\email{alessandro.braggio@nano.cnr.it}
\affiliation{NEST, Istituto Nanoscienze-CNR and Scuola Normale Superiore, Piazza San Silvestro 12, I-56127 Pisa, Italy}
\author{Francesco Giazotto\,\orcidlink{0000-0002-1571-137X}}
\email{francesco.giazotto@nano.cnr.it}
\affiliation{NEST, Istituto Nanoscienze-CNR and Scuola Normale Superiore, Piazza San Silvestro 12, I-56127 Pisa, Italy}
\author{Roberta Citro\,\orcidlink{0000-0002-3896-4759}}
\email{rocitro@unisa.it}
\affiliation{Dipartimento di Fisica ``E.R. Caianiello'', Universit\`a di Salerno, Via Giovanni Paolo II, 132, I-84084 Fisciano (SA), Italy}
\affiliation{INFN, Sezione di Napoli Gruppo Collegato di Salerno, Complesso Universitario di Monte S. Angelo, I-80126 Napoli, Italy}
\affiliation{CNR-SPIN c/o Universit\'a degli Studi di Salerno, I-84084 Fisciano (Sa), Italy}

\begin{abstract}
Thermoelectrical properties are frequently used to characterize the materials and endow the free energy from wasted heat for useful purposes. Here, we show that linear thermoelectric effects in tunnel junctions with Fe-based superconductors, not only address the dominance between particle and hole states, but even provide information about the superconducting order parameter symmetry. In particular, we observe that nodal order parameters present a maximal thermoelectric effect at lower temperatures than for nodeless cases.
Furthermore, we show also that superconducting tunnel junctions between Fe-based and BCS superconductors could provide a thermoelectric efficiency ZT exceeding 6 with a linear Seebeck coefficient around $S\approx 800\;\mu\text{V/K}$ at a few Kelvin. These results pave the way to novel thermoelectric machines based on multi-band superconductors.
\end{abstract}

\maketitle

\emph{Introduction} $-$
Physical systems based on hybrid superconducting junctions have
demonstrated a great potential for energy management issues~\cite{For17,Hwa20,giazotto_rmp_2006,pekola_2012}. Recently, they also attracted interest for their unexpectedly good thermoelectric (TE) performance~\cite{Ozaeta2014,kolenda_2016,Bergeret2018,Hussein2019,Marchegiani2020,Blasi20,Blasi20a,Germanese2022}, finding also a role in different quantum technology applications~\cite{Giazotto2015,Hekkila2018,Paolucci2023}. In a two-terminal system,
a necessary condition for thermoelectricity in the linear
regime, i.e., for a small voltage $\delta V$ and a small temperature bias $\delta T$, is breaking of the particle-hole (PH) symmetry. 
A sizable linear TE effect, much larger than that commonly found in metallic structures, has been recently reported in SC-ferromagnet tunnel junctions (TJs)~\cite{Machon2014,Ozaeta2014,kolenda_2016}, due to the spin-dependent effective breaking of PH symmetry. 

Here, we show a robust linear TE effect in hybrid superconducting junctions with a multiband SC, namely, an Fe-based SC (FeSC). Our results, beyond proving a linear TE effect with unconventional SCs, establish TE phenomena as a potential probe of the superconducting order parameter symmetries.
In fact, one of the central problems for unconventional SCs is the nature of the pairing mechanism, which is tightly connected to the order parameter symmetries~\cite{Goll2006Book,Rei12,Citro2017Book,Ben20}. Pairing symmetry is known to be $d$-wave in cuprates, but is still unresolved for the FeSCs. They are unique among unconventional SCs, since different ordering phenomena are present in a multi-orbital scenario~\cite{mazin_2008, wang_2009,Gua21}.
It was first theoretically predicted that FeAs-based high-temperature SCs have a sign change of the order parameter on the Fermi surface~\cite{mazin_2008}.
Experimental evidence for FeSC seems to be favourable to $s_\pm$ pairing, implying that the electron- and hole-like bands both develop an $s$-wave superconducting state with opposite sign in order parameter~\cite{mazin_2008,kontani_2008}.
Point contact Andreev reflection spectroscopy was also been applied to FeSCs to probe the order parameter symmetry~\cite{chen_2008,daghero_2012}. However, the results of these studies are not completely conclusive due to the complexity 
of the Andreev reflection spectra of a normal metal/multiband-SC interface. 
\begin{figure}[b!!]
\centering
\includegraphics[width=0.9\columnwidth]{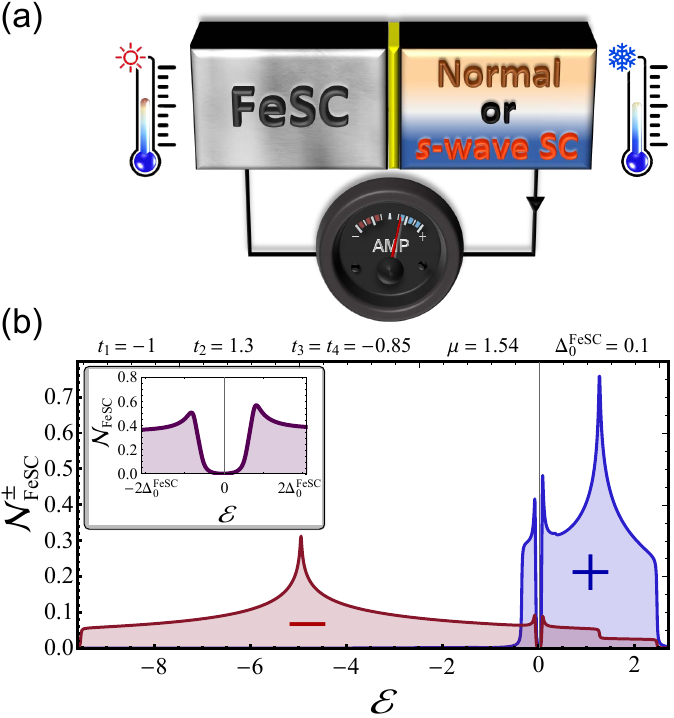}
\caption{(a) Cartoon of the thermally-biased TJ formed by an FeSC and a normal or a $s$-wave superconducting electrode. (b) Electron-like ($+$, blue curve) and hole-like ($-$, red curve) band contributions to the DoS, $\mathcal{N}_{\rm{FeSC}}^\pm \left ( \varepsilon \right )$, see Eq.~\eqref{eq:IBDoSpm}. The hopping parameters, in units of $\left| t_1\right|$, are $(t_1, t_2, t_3, t_4) = (-1, 1.3,-0.85,-0.85)$. The chemical potential is taken at half filling $\mu=1.54$, and the gap size is assumed to be $\Delta_0^{\text{FeSC}}=0.1$. The inset shows a magnifications of the total DoS, $\mathcal{N}_{\rm{FeSC}} \left ( \varepsilon \right )$, in the energy range $\left|\varepsilon \right|\leq 2\Delta_0^{\small\text{FeSC}}$.
}
\label{Figure01}
\end{figure}
Notably, bulk thermoelectrical measurements of FeSC have been reported for the normal state~\cite{Pallecchi2016}, although the analysis is quite intricate due to the competition of different mechanisms, such as phonon-~\cite{Ziman2001} and magnon-drag phenomena~\cite{Caglieris2014}, in a multiband setting. The linear TE properties of a TJ between a FeSC and a normal metal allow identifying the strong intrinsic PH asymmetry in the FeSC density of states (DoS) in a quite direct way, \emph{also for} the superconducting phase. We will also potentially discriminate between different order parameter symmetries. Furthermore, if the normal-metal is replaced by a Bardeen–Cooper–Schrieffer (BCS) SC, we observe also astounding TE figures of merit, which may be relevant for energy harvesting applications and quantum technologies~\cite{Hekkila2018,Patent1,Patent2,Paolucci2023}. 

\emph{Thermoelectrical transport} $-$ In order to address the physics of the FeSC, we focus on the linear TE properties of a TJ between FeSC and a normal metal (or a SC), see Fig.~\ref{Figure01}(a). 
The linear-response coefficients of charge ($I$) and heat ($\dot{Q}$) currents can be expressed in terms of the Onsager matrix~\cite{Ben17}
\begin{equation}
\begin{pmatrix}
I \\
\dot{Q}
\end{pmatrix}=\begin{pmatrix}
\sigma & \alpha \\
\alpha & \kappa T \\
\end{pmatrix}\begin{pmatrix}
\delta V \\
\delta T/T
\end{pmatrix},
\end{equation}
where we assumed time-reversal symmetry satisfied and a small voltage (temperature) bias $\delta V$ ($\delta T$).
Here, $\alpha$ is the TE coefficient, while $\sigma$ and $\kappa$ are the electric and thermal conductances, respectively. At the lowest order of tunneling, it is easy to express those linear coefficients as
\begin{equation}\label{alpha}
\begin{pmatrix}
\sigma \\
\alpha \\
\kappa 
\end{pmatrix}=\frac{G_T}{e}\!\bigintss_{-\infty}^{\infty}\begin{pmatrix}
e \\
\varepsilon \\
\varepsilon^2\big/eT
\end{pmatrix}\frac{\mathcal{N}_L\left ( \varepsilon\right )\mathcal{N}_R\left ( \varepsilon\right )d\varepsilon}{4k_{\text{B}}T\cosh^2\!\left ( \varepsilon/2k_{\text{B}}T \right )}
\end{equation}
in terms of the lead DoS, $\mathcal{N}_j\left ( \epsilon \right )$ with $j=L,R$, where $G_T$ is the normal-state conductance of the junction, $-e$ is the electron charge, and $k_{\text{B}}$ is the Boltzmann constant. 
In our analysis, we take into account two different cases, i.e., a junction formed between a FeSC 
tunnel coupled with a normal lead, i.e., with $\mathcal{N}_R(\epsilon)=1$ being the energy-independent normalized DoS  in Eq.~\eqref{alpha}, or alternatively with an $s$-wave SC, with $\mathcal{N}_R\left ( \epsilon \right )=\left | \text{Re}\left [ \frac{ \epsilon +i\Gamma}{\sqrt{(\epsilon +i\Gamma) ^2-\Delta^2\left ( T \right )}} \right ] \right |$, where $\Gamma=\gamma\Delta_0$ is the phenomenological Dynes parameter~\cite{Dyn78}, $\Delta_0=1.764 k_{\text{B}}T_{c}$, and $T_{c}$ is the critical temperature of the BCS SC. In this case, it is also implicitly assumed that the Josephson coupling~\cite{GuaBra19,Gua19} between the two superconducting leads is strongly suppressed~\cite{Note1}. 
However, let us point out that most of the results shown in the following, particularly those in which we will look at the effect of the order parameter symmetry on the TE response, are obtained by considering an FeSC-I-N junction, where the Josephson effect is not even present.

To calculate the DoS of FeSC we rely on the two-orbital, four-band tight-binding approach by Raghu \emph{et.~al}~\cite{Rag08}, which is the minimal model~\cite{Note2}. 
 The diagonalization of this tight-binding Hamiltonian model leads to eigenvalues that can be written in a compact form as
\begin{equation}\label{eigenvalues}
\varepsilon_{\mathbf{k}\pm}^d=\sqrt{\left(\varepsilon_{\mathbf{k}\pm}^0\right)^2+\left|\vec{d}_{\mathbf{k},g} \right|^2}.
\end{equation}
Here $\varepsilon_{\mathbf{k}\pm}^0=\xi_{\mathbf{k}+}-\mu\pm\sqrt{\xi^2_{\mathbf{k}xy}+\xi^2_{\mathbf{k}-}}$, $\mu$ is the chemical potential, and $\vec{d}_{\mathbf{k},g}=\big ( 0,0,s_{x^2y^2} \big )$, where $s_{x^2y^2}\equiv s_\pm=\Delta_0^{\text{FeSC}}\cos k_x\cos k_y$, $\Delta_0^{\text{FeSC}}$ is the gap size, and $\left|\vec{d}_{\mathbf{k}} \right|^2$ represents the effective amplitude of the pairing interactions~\cite{Par08}. For the explicit expressions of $\xi_{\mathbf{k}\pm}$ and $\xi_{\mathbf{k}xy}$ we refer the reader to Ref~\cite{NoteSM}. We adopted the $s_\pm$-wave state, which is the mostly accepted FeSC pairing state~\cite{Ban17,Fer22}. However, the multiband character of FeSCs offers also chances for more exotic pairing states~\cite{Par08,Fer22}, and thus we will also discuss TE properties with other order parameter symmetries in the following. Hereafter, we took the interorbital hopping parameter $\left| t_1 \right|$ as a standard unit of energy, and temperature will be measured in units of $\left| t_1 \right|/k_B$~\cite{Note3}. 
%
\begin{figure*}[t!!]
\centering
\includegraphics[width=2\columnwidth]{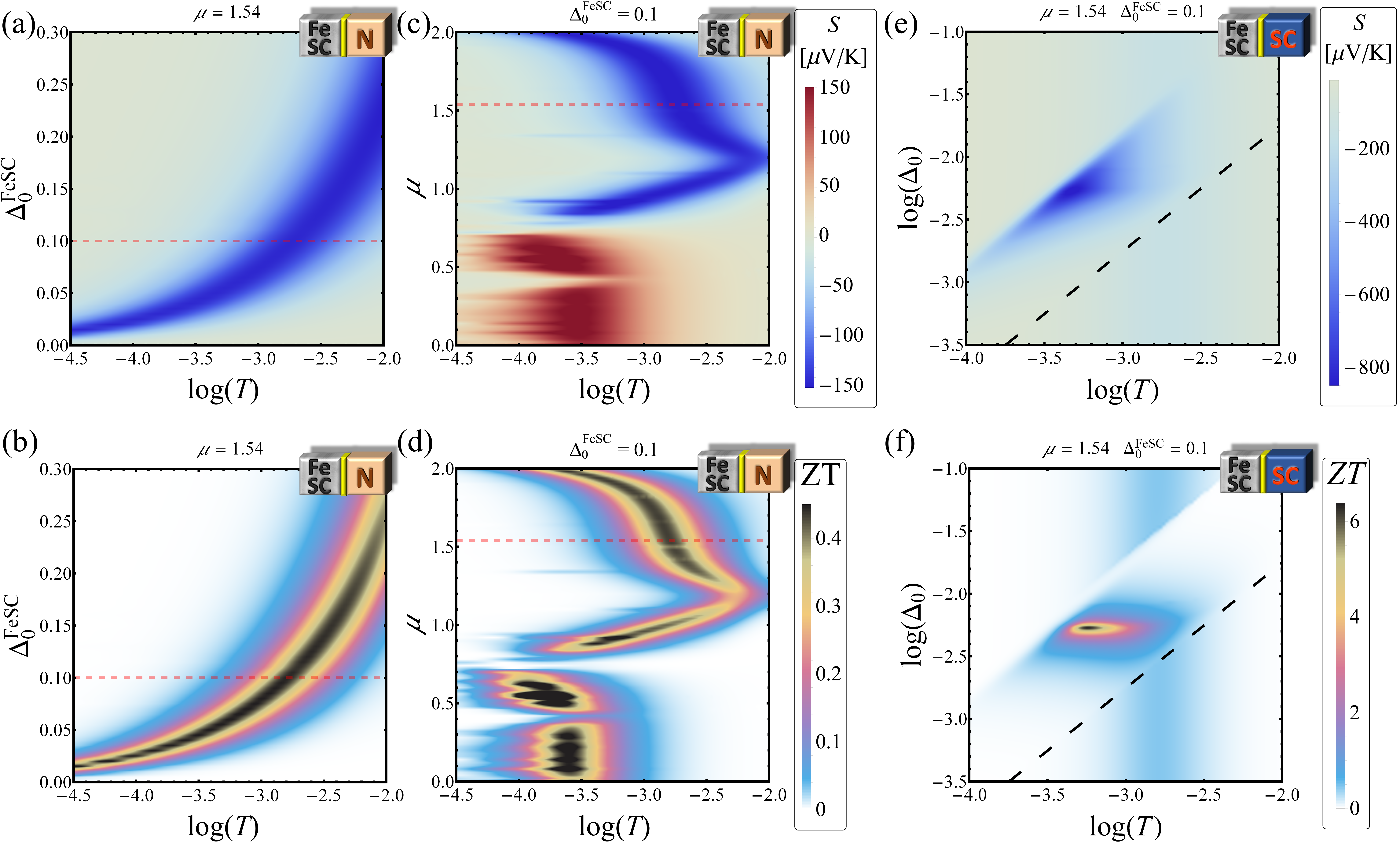}
\caption{FeSC-I-N junction: (a) and (b) Seebeck coefficient, $S(T,\Delta_0^{\text{FeSC}})$, and figure of merit, ZT$(T,\Delta_0^{\text{FeSC}})$ at $\mu=1.54$. (c) and (d) $S(T,\mu)$ and ZT$(T,\mu)$, at $\Delta_0^{\text{FeSC}}=0.1$. The legends in panels (c) and (d) refer also to panels (a) and (b), respectively, and the red dashed line marks the condition $(\Delta_0^{\text{FeSC}},\mu)=(0.1,1.54)$. FeSC-I-S junction: (e) and (f) $S(T,\Delta_0)$ and ZT$(T,\Delta_0)$, at $\mu=1.54$, $\Delta_0^{\text{FeSC}}=0.1$, and $\gamma=10^{-4}$. The dashed lines in panels (c) and (f) mark the values $\Delta^{th}_0=1.764T$. A cartoon in the top-right corner of each panel helps to recognise the type of junction considered at a glance.}
\label{Figure02}
\end{figure*}
Finally, the FeSC total DoS turns out to be the sum of an electron-like [$+$, blue curve in Fig.~\ref{Figure01}(b)] and a hole-like [$-$, red curve in Fig.~\ref{Figure01}(b)] band contribution~\cite{Pto14}, $\mathcal{N}_{\rm{FeSC}} \left ( \varepsilon \right )=\mathcal{N}_{\rm{FeSC}}^+ \left ( \varepsilon \right )+\mathcal{N}_{\rm{FeSC}}^- \left ( \varepsilon \right )$, where
\begin{equation}\label{eq:IBDoSpm} 
\mathcal{N}_{\rm{FeSC}}^\pm \left ( \varepsilon \right )\!=\!\sum_{\mathbf{k}}\frac{\varepsilon + \varepsilon_{\mathbf{k}\pm}^0}{2 \varepsilon_{\mathbf{k}\pm}^d}\left\{\delta \left [ \varepsilon_{\mathbf{k}\pm}^d-\varepsilon \right ]-\delta \left [ \varepsilon_{\mathbf{k}\pm}^d+\varepsilon \right ] \right\}\!.\!\!\!
\end{equation}
Note also that the superconducting instability opens the gap symmetrically around the chemical potential, as illustrated in the inset of Fig.~\ref{Figure01}(b) (see also Ref~\cite{NoteSM}).

\emph{Figures of merit} $-$ In order to quantify the TE performance, it is usual to consider the Seebeck coefficient $S=-\alpha/(\sigma T)$ and thermodynamic efficiency with the dimensionless figure of merit $\text{ZT}=S^2\sigma T/\left [ \kappa-\alpha^2/(\sigma T)\right ]$~\cite{Note3.5}. A large value of ZT means better thermodynamic efficiency and if it tends to infinity, the efficiency of the device tends to the Carnot limit. In the following, we show how $S$ and ZT of an FeSC-insulator-normal metal (FeSC-I-N) junction depend on the temperature, considering possible different values of $\Delta_0^{\text{FeSC}}$~\cite{Note4} 
and changing the doping level, $\mu$~\cite{Note5}.

Figures~\ref{Figure02}(a) and (b) collect the $S(T,\Delta_0^{\text{FeSC}})$ and ZT$(T,\Delta_0^{\text{FeSC}})$ maps at the half-filling condition $\mu= 1.54$. For a given $\Delta_0^{\text{FeSC}}$, both the Seebeck coefficient and the TE efficiency behave non-monotonically, with a clear maximum that shifts towards gradually higher temperatures as $\Delta_0^{\text{FeSC}}$ increases. Indeed,  $\Delta_0^{\text{FeSC}}$ is the energy scale that mainly influences the optimal temperature that maximizes the TE effect. Furthermore, by increasing $\Delta_0^{\text{FeSC}}$ the subgap states reduce, correspondingly requiring higher energies to achieve the same TE effect. The use of FeSCs is a further advantage over conventional SC configurations, since it allows operation at higher temperatures that provide a larger Seebeck coefficient.

Figures~\ref{Figure02}(c) and (d) show what happens when changing the doping level, keeping fixed $\Delta_0^{\text{FeSC}}=0.1$~\cite{Note6}. 
The FeSC DoS dependence on $\mu$ is illustrated in Ref~\cite{NoteSM}. The $S(T,\mu)$ map still reveals single-peaked profiles, but modifying the doping we observe the inversion of the Seebeck coefficient sign around $\mu\sim0.75$, below (above) which $S>0$ ($S<0$) for the hole (electron) DoS contribution dominates. We note that the point at which the Seebeck coefficient changes sign differs from the half-filling condition.
The reason of that is the lack of symmetry between particle-like and hole-like bands for an FeSC in the energy window determined by the working temperatures.

In the cases discussed so far, we achieve Seebeck coefficients up to $\left| S \right|\sim 150~\mu\text{V/K}$ reaching also TE efficiencies of $\text{ZT}\sim 0.5$. In the case of an undoped FeSC, $\mu=1.54$, with $\Delta_0^{\text{FeSC}}=0.1$, i.e., the red dashed lines in Fig.~\ref{Figure02}(a-d), the highest TE efficiency is reached at $T\simeq1.6\times10^{-3}$, which may be a temperature around $2.8\;\text{K}$.

We remark that the maximum Seebeck coefficient we obtain is several orders of magnitude larger than that usually found in metallic structures at the same temperatures. However, the thermoelectricity of the FeSC-I-N junction outperforms magnetic TJs~\cite{Wal11} and is quite well comparable with hybrid SCs-ferromagnets TJs~\cite{Kol16,Gon23} and quantum-dot setup~\cite{Svi16}.\\
It is noteworthy to show that, if we replace the normal metal with a BCS SC with a gap $\Delta_0$, 
the linear thermoelectricity can be further enhanced. This effect can be ascribed to the additional contribution of the conventional DoS peaks intertwined with the multi-band character of the FeSC.
Thus, in Fig.~\ref{Figure02}(e) and (f) we present the $S(T,\Delta_0)$ and ZT$(T,\Delta_0)$ maps of a FeSC-insulator-SC (FeSC-I-S) TJ. In this case, we assume a specific FeSC with a given gap $\Delta_0^{\text{FeSC}}=0.1$ and we explore the TE response at different values of the BCS superconducting gap $\Delta_0$. A region of the $(T,\Delta_0)$ parameter space emerges in which both Seebeck coefficient and TE efficiency increase significantly, even reaching the values $\left| S \right|\sim 870~\mu\text{V/K}$ and ZT$\sim6.5$ at $(T,\Delta_0)_{max}\simeq(0.63,5.3)\times10^{-3}$. In natural units, these quantities correspond to $T\simeq1.8\;\text{K}$ for a BCS SC with $T_c\simeq5.2\;\text{K}$~\cite{Note7}. %

%
\begin{figure}[t!!]
\centering
\includegraphics[width=\columnwidth]{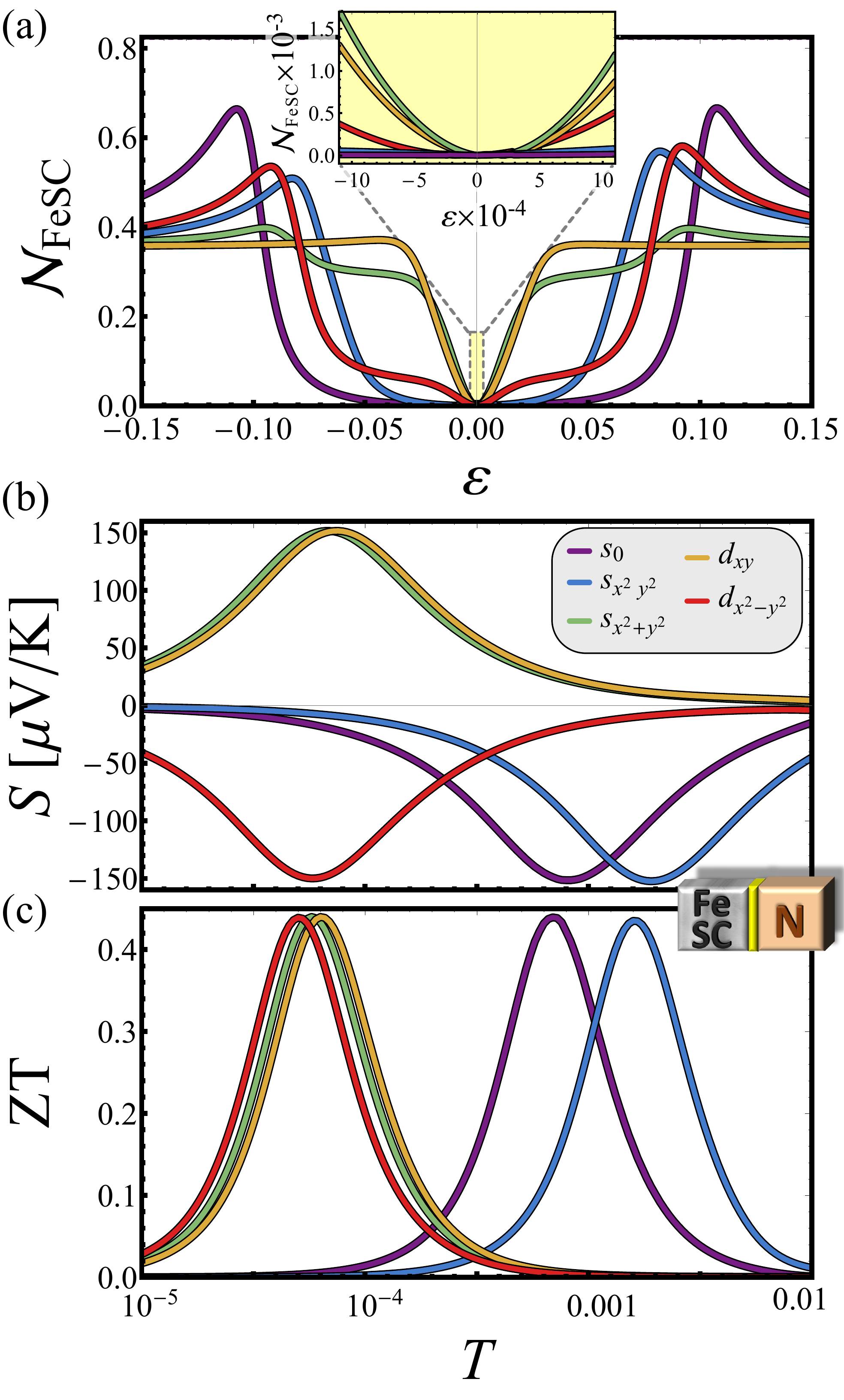}
\caption{FeSC-I-N junction: (a) DoS, $\mathcal{N}_{\rm{FeSC}} \left ( \varepsilon \right )$, (b) Seebeck coefficient, $S(T)$, and (c) figure of merit, ZT$(T)$ by changing the symmetry of the order parameter, at $\mu=1.54$ and $\Delta_0^{\text{FeSC}}=0.1$. The inset in (a) illustrates the low-energy behaviour of the DoSs. The legend in panel (b) refers to all panels.}
\label{Figure03}
\end{figure}
%

\emph{Order parameter symmetry detection} $-$ We show here that TE figures of merit are also a powerful tool for addressing order-parameter symmetry (OPS). We compare the temperature dependence of $S$ and ZT of an FeSC-I-N TJ, with $\mu=1.54$ and $\Delta_0^{\text{FeSC}}=0.1$, taking into account different OPSs: we cover the three possible $s$-wave symmetries, i.e., the constant-gap case $s_0$, $s_{x^2y^2}=\Delta_0^{\text{FeSC}}\cos k_x\cos k_y$, and $s_{x^2+y^2}=\Delta_0^{\text{FeSC}}(\cos k_x+\cos k_y)/2$, and the two $d$-wave symmetries, i.e., $d_{xy}=\Delta_0^{\text{FeSC}}\sin k_x\sin k_y$ and $d_{x^2-y^2}=\Delta_0^{\text{FeSC}}(\cos k_x-\cos k_y)/2$~\cite{Note8}. 
Since it will be useful later on, we recall that the $d_{x^2-y^2}$, $d_{xy}$, and $s_{x^2+y^2}$ OPSs are nodal, while the others are nodeless~\cite{Par08,Seo08}.

Figure~\ref{Figure03} demonstrates that the TE figures of merit can provide valuable clues for determining the OPS of the system. Starting from the top of this figure, panel (a) takes a closer look at the different DoSs in play, with the inset serving to highlight the low-energy region. Figures~\ref{Figure03}(b)-(c) illustrate the Seebeck coefficient $S(T)$ and the TE efficiency ZT$(T)$, both showing, in a semilog scale, bell-shaped, single-peaked profiles for each symmetry considered. It immediately stands out that the ``position'' of these peaks depends strongly on the OPS: indeed, for the $s_0$ and $s_{x^2y^2}$ cases, both $S$ and ZT are peaked roughly around $T^{\rm peak}\sim10^{-3}$, while the other symmetries give $S$ and ZT peaks centred on temperatures an order of magnitude lower. To give realistic numbers (here the subscript distinguishes symmetries), the $S$ peak for nodeless OPS is located at $T_{x^2y^2}^{\rm peak}\simeq3.3\;\text{K}$(blue), $T_{0}^{\rm peak}\simeq1.4\;\text{K}$(violet), whereas for nodal cases one finds $T_{xy}^{\rm peak}\simeq0.13\;\text{K}$(yellow), $T_{x^2+y^2}^{\rm peak}\simeq0.12\;\text{K}$(green), and $T_{x^2-y^2}^{\rm peak}\simeq0.10\;\text{K}$(red). 
To grasp this result, we recall that the energy window relevant for calculating the TE coefficients scales with temperature, i.e., $\left| \varepsilon \right|\sim T$~\cite{NoteSM}. 
For instance, the energies considered in the inset of Fig.~\ref{Figure03}(a) are essentially those where one should focus if $T\sim10^{-4}$. Here, it is evident that only the $d_{xy}$, $s_{x^2+y^2}$ and $d_{x^2-y^2}$ DoSs (i.e., those showing ZTs peaked at these temperatures) are clearly non-zero. We see that nodeless FeSC pairings present maximal thermoelectricity (absolute value of the Seebeck coefficient) at relatively high temperatures. Instead, for the nodal cases, due to the presence of low energy states in the gap, the maximal thermoelectricity is observed at much lower temperature regimes. This result is quite robust against variations of the gap amplitude, hopping parameter, and chemical potential values~\cite{NoteSM}.
Furthermore, we also see that the first two DoSs, i.e., the yellow and green curves, are unbalanced toward the hole side ($\varepsilon<0$), unlike the third, i.e., the red curve, which appears to be slightly unbalanced toward particle side ($\varepsilon>0$). This is directly reflected in the sign of $S$, which immediately tells us the PH asymmetry of the FeSC DoS and, therefore, $S>0$ ($S<0$) in the former (latter) case. This is clearly confirmed also by looking the PH asymmetry of the DoSs on the larger energy scale considered in Fig.~\ref{Figure03}(a). It is evident, for instance, that the $s_0$ and $s_{x^2y^2}$ OPSs, i.e., the violet and blue curves, respectively, are unbalanced to the right, i.e., the particle contribution dominates, thereby making $S$ negative.

We emphasize that, in principle, the measurement of the PH asymmetry could be addressed by directly measuring the tunneling differential conductance. However, a systematic experimental asymmetry in the bias polarization of the junction cannot be easily excluded loosing sensitivity for small PH asymmetry. The linear thermoelectricity much safely returns this information in an independent way. Yet, we stress that the thermoelectrical signatures discussed in this Letter, being associated with the quasiparticle tunneling in the junction, are not affected by any phonon- or magnon-drag effects, which instead usually influence the bulk TE properties in the normal phase~\cite{Pallecchi2016}. 

\emph{Conclusions} $-$ To summarize, we have demonstrated that an FeSC TJ can show sizable TE efficiency, and that both the TE figure of merit and the Seebeck coefficient are found to be non-monotonic, single-peaked functions of temperature. Moreover, they can provide details on the underlying symmetry of the order parameter addressing the PH asymmetry of the DoS. In particular, we demonstrated in FeSC-I-N junction that the position of both the ZT and $S$ peaks allows us to clearly distinguish nodal from nodeless symmetries. Furthermore, the sign of $S$ provides further information on the asymmetry of PH, distinguishing cases where the TE efficiency is not discriminating, such as for the two $d$-waves symmetries. Our results also establish the relevance of multiband SCs for a novel generation of TE devices.

As a closing remark, we observe that the proposed approach may be used for studying other quantum materials~\cite{Fer22}.
Multi-orbital pairing approaches have been widely used also to shed light in other multiband SCs, such as ruthenates and nickelates, and insights about Hund-metal. 
Therefore, the TE-based investigation of tunneling junctions presented in this Letter complements the actual experimental techniques, finding a fertile ground for the study of novel quantum materials.\\

\begin{acknowledgments}
F.G. acknowledges PNRR MUR project PE0000023-NQSTI for partial financial support. F.G and A.B. acknowledge the EU’s Horizon 2020 research and innovation program under Grant Agreement No. 964398 (SUPERGATE) and No. 101057977 (SPECTRUM) for partial financial support. 
A.B. acknowledges the Royal Society through the International Exchanges between the UK and Italy (Grants No. IEC R2 212041).
\end{acknowledgments}

\bibliographystyle{apsrev4-1}

%

\widetext
\clearpage
\begin{center}
\textbf{\large Supplemental Material: Thermoelectric signatures of order-parameter symmetries in iron-based superconducting tunnel junctions}
\end{center}
\setcounter{equation}{0}
\setcounter{figure}{0}
\setcounter{table}{0}
\setcounter{page}{1}
\setcounter{section}{0}
\makeatletter
\renewcommand{\thetable}{S\arabic{table}}
\renewcommand{\thesection}{S-\Roman{section}}
\renewcommand{\theequation}{S\arabic{equation}}
\renewcommand{\thefigure}{S\arabic{figure}}
\renewcommand{\bibnumfmt}[1]{[S#1]}
\renewcommand{\citenumfont}[1]{S#1}

\subsection{Model for an FeSC}
\label{AppB}

\begin{figure*}[t!!]
\centering
\includegraphics[width=\columnwidth]{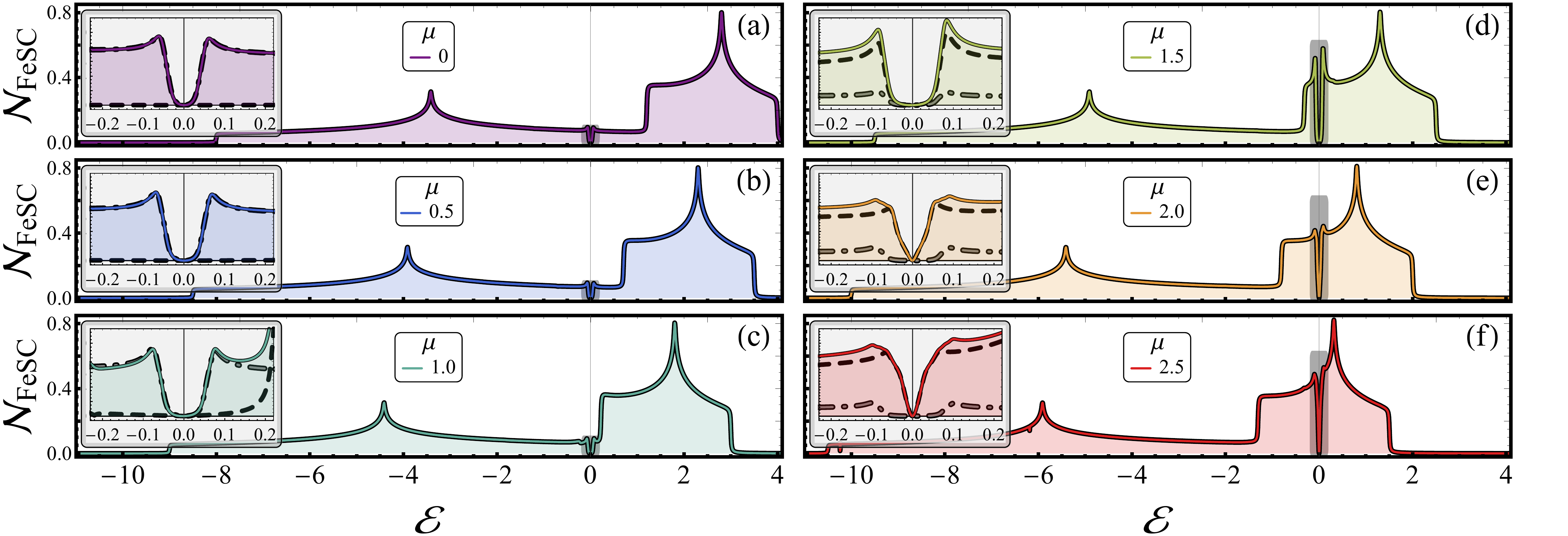}
\caption{DoSs, $\mathcal{N}_{\rm{FeSC}} \left ( \varepsilon \right )$, at different values of $\mu\in[0-2.5]$, $\Delta_0^{\text{FeSC}}=0.1$, and $(t_1,t_2,t_3,t_4) = (-1,1.3,-0.85-0.85)$. The insets sketch a zoom of both $\mathcal{N}_{\rm{FeSC}} \left ( \varepsilon \right )$ and $\mathcal{N}^\pm_{\rm{FeSC}} \left ( \varepsilon \right )$, see the solid, dashed ($+$), and dot-dashed lines ($-$), respectively, in the energy range $\left|\varepsilon \right|\lesssim 2 \Delta_0^{\small\text{FeSC}}$. which is highlighted by a gray shaded area in the main plot.}
\label{FigureSM04}
\end{figure*}

We use the two-orbital, four-band tight-binding model developed by Raghu \emph{et.~al}~\cite{Rag08} (see also Refs.~\cite{Par08,Seo08} for more details). The diagonalization of this tight-binding Hamiltonian model leads to eigenvalues that can be written in a compact form as
\begin{equation}\label{eigenvalues}
\varepsilon_{\mathbf{k}\pm}^d=\sqrt{\left(\varepsilon_{\mathbf{k}\pm}^0\right)^2+\left|\vec{d}_{\mathbf{k},g} \right|^2}\qquad\qquad\text{where}\qquad\qquad
\varepsilon_{\mathbf{k}\pm}^0=\xi_{\mathbf{k}+}-\mu\pm\sqrt{\xi^2_{\mathbf{k}xy}+\xi^2_{\mathbf{k}-}}.
\end{equation}
Here $\mu$ is the chemical potential and $\vec{d}_{\mathbf{k},g}=\big ( 0,0,s_{x^2y^2} \big )$, where $s_{x^2y^2}=\Delta_0^{\text{FeSC}}\cos k_x\cos k_y$, $\Delta_0^{\text{FeSC}}$ is the gap size.
In Eq.~\eqref{eigenvalues}, we take into account the kinetic energy terms $\xi_{\mathbf{k}\alpha\beta}$ of a particle with momentum $\mathbf{k}$ changing the orbital from $\beta$ to $\alpha$, given by 
\begin{eqnarray}
\xi_{\mathbf{k}\pm}&\!=\!&(\xi_{\mathbf{k}xx}\pm\xi_{\mathbf{k}yy})/2\\\nonumber
\xi_{\mathbf{k}xx}&\!=\!&-2t_1 \cos (k_x)\!-\!2t_2 \cos (k_y)\!-\!4 t_3 \cos (k_x) \cos (k_y),\\\nonumber
\xi_{\mathbf{k}yy}&\!=\!&-2t_2 \cos (k_x)\!-\!2t_1 \cos (k_y)\!-\!4 t_3 \cos (k_x) \cos (k_y),\\\nonumber
\xi_{\mathbf{k}xy}&\!=\!&-4 t_4 \sin (k_x) \sin (k_y).
\end{eqnarray} 
The coefficients $t_1$ and $t_2$ are the intraorbital nearest-neighbor hopping amplitudes, while $t_3$ and $t_4$ are the intraorbital and interorbital next-nearest-neighbor hopping amplitudes, respectively~\cite{Liu18}. Different sets of hopping parameters correspond to different Fe-based superconductors (FeSC) (as in the manuscript, energies and temperatures are in units of $\left| t_1 \right|$ and $\left| t_1 \right|/k_B$, respectively), i.e., $(t_1,t_2,t_3,t_4) = (-1,1.3,-0.85-0.85)$ for iron pnictdes~\cite{Rag08}, or $(t_1,t_2,t_3,t_4) = (-1,1.5,-1.2-0.95)$ for iron selenides~\cite{Yam13,Dum16}. In this article, we focus primarily on iron pnictdes, although, as we show in the following, the main claims about the capability of thermoelectric (TE) measurements to resolve the pairing symmetry remain valid even using the hopping parameter values for selenides (even if for this kind of FeSC more sophisticated modelling has proven to be more effective~\cite{Nic17}). 

\subsection{Density of states of a FeSC}
For the sake of convenience, we rewrite the FeSC density of states (DoS) presented in the main text as
\begin{equation}\label{eq:IBDoSpm} 
\mathcal{N}_{\rm{FeSC}}^\pm \left ( \varepsilon \right )=\frac{\eta}{\pi^3} \iint_0^{\pi}dk_xdk_y\frac{\varepsilon + \varepsilon_{\mathbf{k}\pm}^0}{2 \varepsilon_{\mathbf{k}\pm}^d}
\left [ \frac{1}{\left (\varepsilon - \varepsilon_{\mathbf{k}\pm}^d \right )^{\!2}+\eta^2}
-\frac{1}{\left (\varepsilon + \varepsilon_{\mathbf{k}\pm}^d \right )^{\!2}+\eta^2} \right ],
\end{equation}
assuming a Lorentzian energy broadening like i.e., $\delta\left ( \varepsilon -\varepsilon_{\mathbf{k}\pm}^{d} \right )=\eta\Big/ \left\{\pi \left [ \left ( \varepsilon -\varepsilon_{\mathbf{k}\pm}^{d} \right )^{2}+\eta^2 \right ] \right\} $~\cite{Par08} (we set $\eta=0.01$ in line with Ref.~\cite{Pto20}).\\
\begin{figure}[t!!]
\centering
\includegraphics[width=0.51\columnwidth]{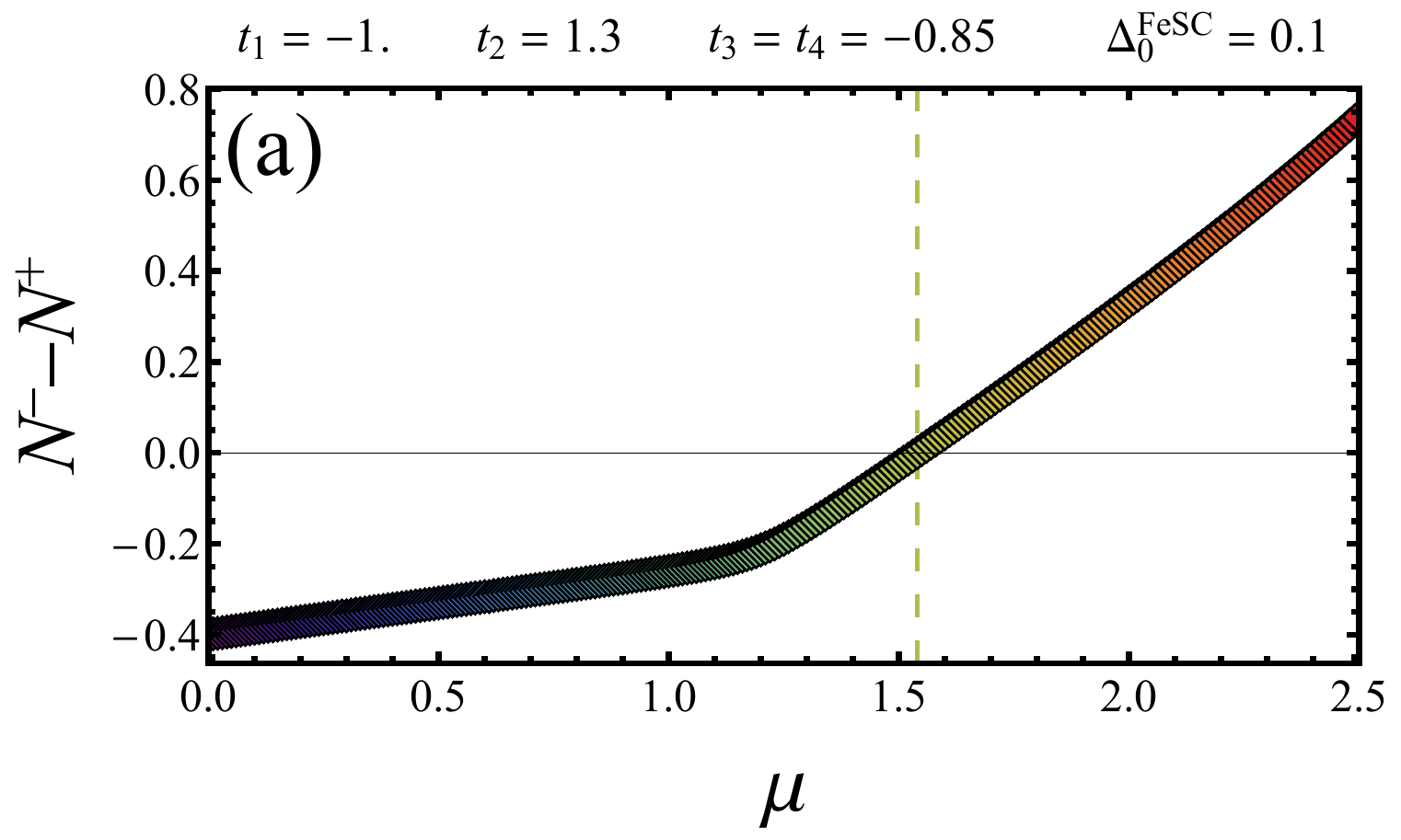}
\includegraphics[width=0.475\columnwidth]{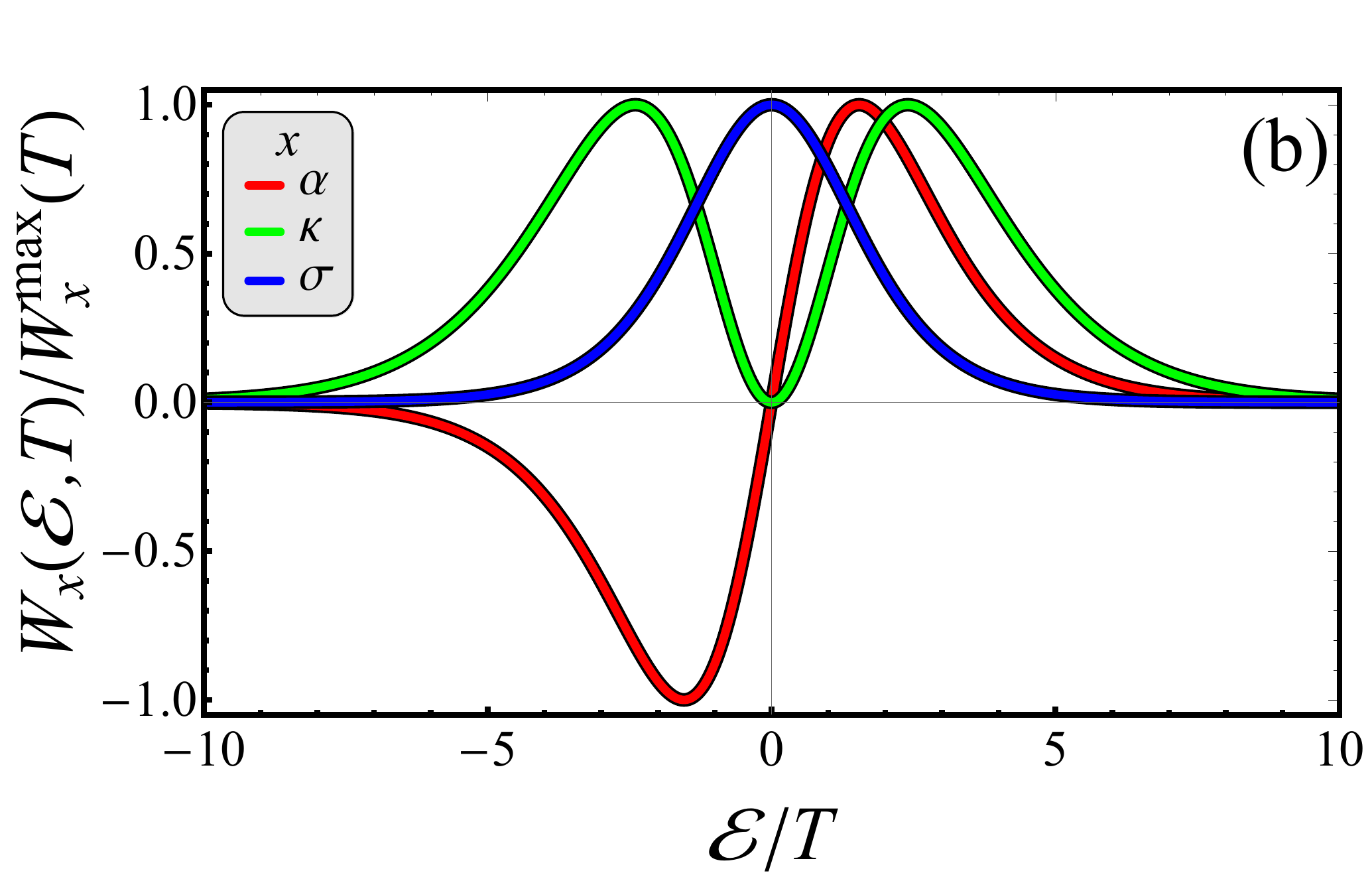}
\caption{(a) $N^-(\mu)-N^+(\mu)$, where $N^{\pm}(\mu)=\pm\int_{0}^{\pm\infty}\mathcal{N}_{\rm{FeSC}} \left ( \varepsilon,\mu \right )d\varepsilon$, as a function of $\mu$. (b) Normalized weight functions, $W_x\left ( \varepsilon, T\right )$, in Eqs.~\eqref{alpha1}-\eqref{alpha2} as a function of $\varepsilon/T$.}
\label{FigureSM02}
\end{figure}
The behavior of the DoS, $\mathcal{N}_{\rm{FeSC}} \left ( \varepsilon \right )$, at different values of $\mu\in[0-2.5]$ and $\Delta_0^{\text{FeSC}}=0.1$ is shown in Fig.~\ref{FigureSM04}. We clearly observe that by changing the doping level the DoS shifts, but leaving the gap opening around $\varepsilon=0$. Moreover, as the doping increases, the DoSs gradually change from fully gap to gapless behaviour, as expected according to Ref.~\cite{Par08} (see insets in Fig.~\ref{FigureSM04} where a zoom is reported); in fact, for a large enough doping one finds a fully gapped hole-like Fermi surface, giving a gap in the DoS at low energies, and a partially gapped electron-like Fermi surface, giving a DOS growing quasilinearly around $\varepsilon=0$. In order to illustrate this behavior, we include in the insets of Fig.~\ref{FigureSM04} the DoSs $\mathcal{N}_{\rm{FeSC}} \left ( \varepsilon \right )$ (solid line), $\mathcal{N}^+_{\rm{FeSC}} \left ( \varepsilon \right )$ (dashed line), and $\mathcal{N}^+_{\rm{FeSC}} \left ( \varepsilon \right )$ (dot-dashed line) within the energy range $\left|\varepsilon \right|\lesssim 2 \Delta_0^{\small\text{FeSC}}$. It is evident that for $\mu=0,$ and $0.5$, see Fig.~\ref{FigureSM04}(a-b), the electron-like contribution does not enter in play at low energies, and in fact the hole DoS contribution dominates (as discussed in the main text) the TE response of the system. For $\mu=1$, see Fig.~\ref{FigureSM04}(c), the electronic component starts to play a role and for $\mu=1.5$, see Fig.~\ref{FigureSM04}(d), both the electrons and holes contributions effectively partake in the overall DoS and appear clearly gapped. The last two panels, for $\mu=2$ and $2.5$, demonstrate that the quasi-linear behaviour at low energies of the DoS is entirely ascribable to the electronic component, the hole being still gapped. The sign of the TE coefficient roughly reconcile with this behaviour since for hole-dominated case it is typically positive where for electron-dominated is more negative. \\ 
It is also interesting to look at the behaviour of the quantity $N^{\pm}(\mu)=\pm\int_{0}^{\pm\infty}\mathcal{N}_{\rm{FeSC}} \left ( \varepsilon ,\mu\right )d\varepsilon$; in particular, in Fig.~\ref{FigureSM02}(a) we focus on the difference $N^{-}(\mu)-N^+(\mu)$ as a function of $\mu$. We recall that the undoped compound, having the half-filled, two electrons per site configuration, requires $\mu= 1.54$, namely, the value largely used in the manuscript, at which $N^{-}-N^+=0$. For $\mu$ lower (higher) than the half-filling value, the positive (negative) energy portion of DoS predominates, being $N^{-}-N^+<0$ ($N^{-}-N^+>0$). A change of slope is evident for the threshold value $\mu^{th}\sim1.15$; this is due to the fact that, increasing the doping level, the electron-like band starts to be strongly occupied (e.g., see right panels of Fig.~\ref{FigureSM04}). This mechanism explains why for $\mu\sim\mu^{th}$ there is a strong change in the "movement" of the TE peak with respect to the doping illustrated in Figs.~2(c) and~(d) of the main text. In particular, we see that for $\mu\gtrsim\mu^{th}$ ($\mu\lesssim\mu^{th}$) the TE peak shifts towards lower (higher) temperatures increasing the electron doping. At lowest doping the thermoelectricity changes sign since it is hole dominated and electronic-like band is almost completely depleted. \\

\subsection{TE coefficients of a tunneling barrier}

\begin{figure*}[t!!]
\centering
\includegraphics[width=\columnwidth]{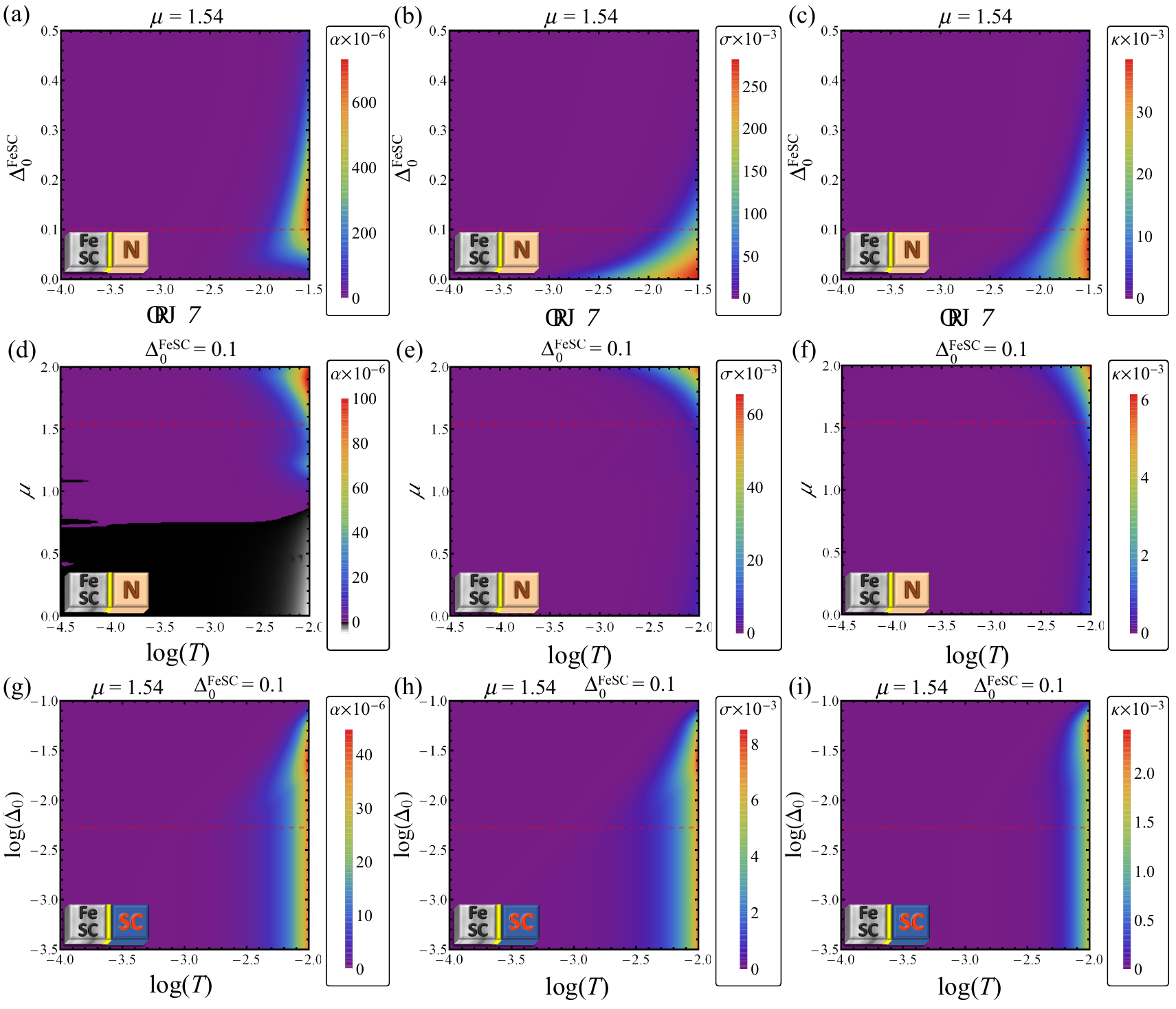}
\caption{FeSC-I-N junction: (a), (b), and (c) $\alpha(T,\Delta_0^{\text{FeSC}})$, $\sigma(T,\Delta_0^{\text{FeSC}})$, and $\kappa(T,\Delta_0^{\text{FeSC}})$ at $\mu=1.54$; (d), (e), and (f) $\alpha(T,\mu)$, $\sigma(T,\mu)$, and $\kappa(T,\mu)$ at $\Delta_0^{\text{FeSC}}=0.1$. FeSC-I-S junction: (g), (h), and (i) $\alpha(T,\Delta_0)$, $\sigma(T,\Delta_0)$, and $\kappa(T,\Delta_0)$ at $\mu=1.54$ and $\Delta_0^{\text{FeSC}}=0.1$. A cartoon in the bottom-left corner of each panel helps to recognise the type of junction considered at a glance.}
\label{FigureSM01}
\end{figure*}

It is useful to express the TE coefficient, $\alpha$, and the electric and the thermal conductances, $\sigma$ and $\kappa$, 
\begin{equation}\label{alpha1}
\begin{pmatrix}
\sigma \\
\alpha \\
\kappa 
\end{pmatrix}=\frac{G_T}{e}\!\bigintss_{-\infty}^{\infty}\begin{pmatrix}
e \\
\varepsilon \\
\varepsilon^2\big/eT
\end{pmatrix}\frac{\mathcal{N}_L\left ( \varepsilon\right )\mathcal{N}_R\left ( \varepsilon\right )d\varepsilon}{4k_{\text{B}}T\cosh^2\!\left ( \varepsilon/2k_{\text{B}}T \right )},
\end{equation}
in such a way as to group all terms excluding DoSs into "weight functions", that is
\begin{equation}
x=\frac{G_T}{e}\int_{-\infty}^{\infty}W_{x}(\varepsilon,T)\mathcal{N}_L ( \varepsilon)\mathcal{N}_R ( \varepsilon)d\varepsilon\qquad\text{with}\qquad x=(\sigma,\alpha,\kappa).\label{alpha2}
\end{equation}
The behavior of the weight functions, $W_x\left ( \varepsilon, T\right )$ with $x=(\sigma,\alpha,\kappa)$, normalised to their maximum value, is shown in Fig.~\ref{FigureSM02}(b). We immediately note that all three functions quickly go to zero for $\left| \varepsilon\right|\gtrsim 10\;T$, thus limiting the temperature range relevant for the calculation of linear TE coefficients. We also observe that $\sigma$ and $\kappa$ are even functions of energy, unlike $\alpha$, which is odd; from this characteristic derives the dependence of the sign of the Seebeck coefficient, $S=-\alpha/(\sigma T)$, on the predominant particle/hole contribution.\\

In Fig.~\ref{FigureSM01} we show how TE coefficient, $\alpha$, and the electric and the thermal conductances, $\sigma$ and $\kappa$, vary in the case discussed in Fig.~2 of the main text (note, in Fig.~\ref{FigureSM01} we are showing dimensionless normalised quantities). In particular: panels (a), (b), and (c) show $\alpha(T,\Delta_0^{\text{FeSC}})$, $\sigma(T,\Delta_0^{\text{FeSC}})$, and $\kappa(T,\Delta_0^{\text{FeSC}})$ at $\mu=1.54$, while panels (d), (e), and (f) show $\alpha(T,\mu)$, $\sigma(T,\mu)$, and $\kappa(T,\mu)$ at $\Delta_0^{\text{FeSC}}=0.1$, in the case of a FeSC-I-N junction. Instead, panels (g), (h), and (i) $\alpha(T,\Delta_0)$, $\sigma(T,\Delta_0)$, and $\kappa(T,\Delta_0)$ at $\mu=1.54$ and $\Delta_0^{\text{FeSC}}=0.1$, in the case of a FeSC-I-S junction. This figure allows us to definitively state that, taken individually, these quantities are unable to provide the information that instead emerges clearly when they are recombined to form the Seebeck coefficient, $S$, and figure of merit, ZT, as shown in Fig.~2 of the article. Finally, we observe also that the black region in Fig.~\ref{FigureSM01}(d) indicates negative $\alpha$ values, giving $S>0$.\\

\subsection{TE and other FeSC order parameter symmetries}
\begin{figure*}[t!!]
\centering
\includegraphics[width=\columnwidth]{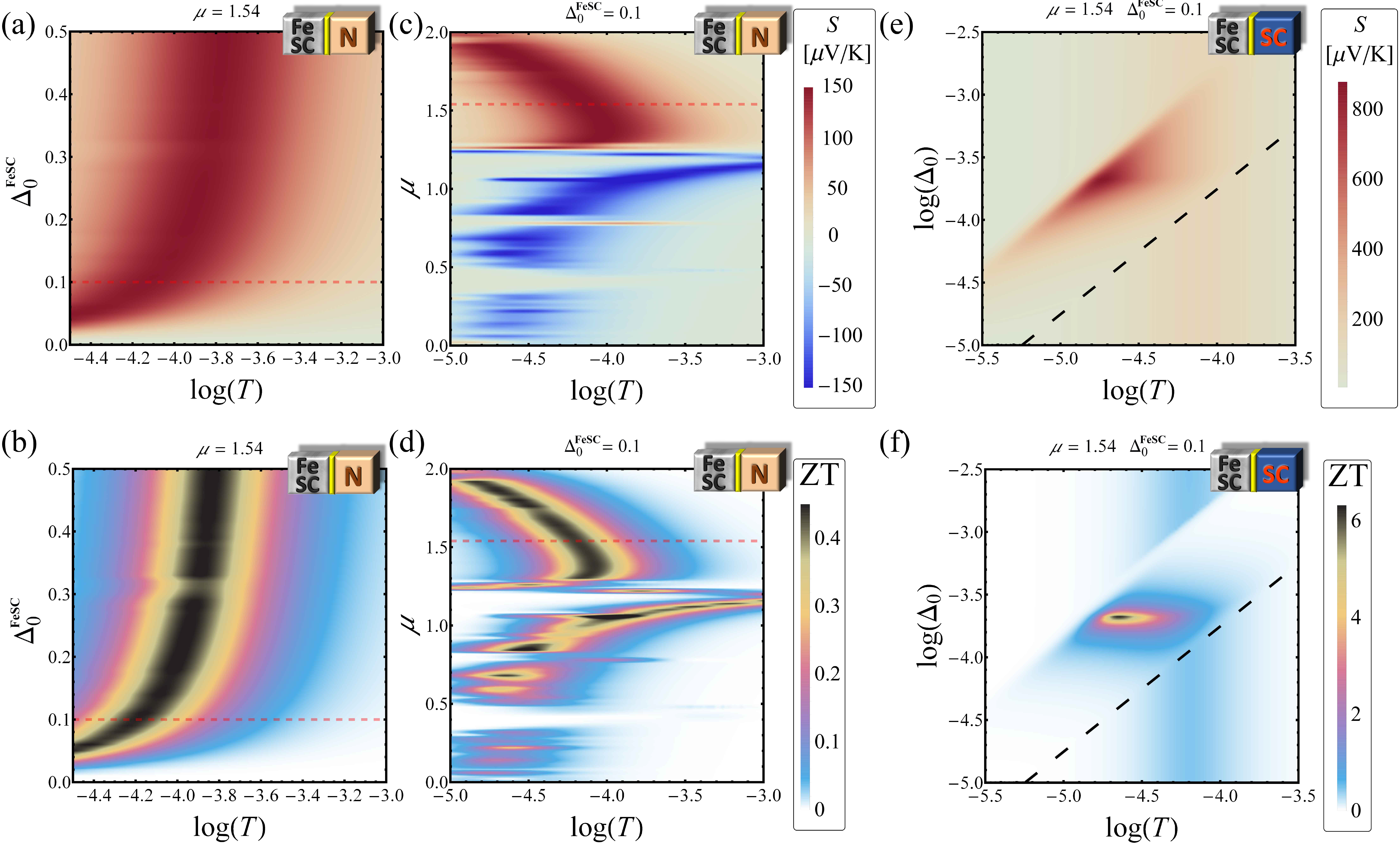}
\caption{TE efficiency in the case of $d_{xy}=\Delta_0^{\text{FeSC}}\sin k_x\sin k_y$ pairing symmetry. FeSC-I-N junction: (a) and (b) Seebeck coefficient, $S(T,\Delta_0^{\text{FeSC}})$, and figure of merit, ZT$(T,\Delta_0^{\text{FeSC}})$ at $\mu=1.54$. (c) and (d) $S(T,\mu)$ and ZT$(T,\mu)$, at $\Delta_0^{\text{FeSC}}=0.1$. The legends in panels (c) and (d) refer also to panels (a) and (b), respectively, and the red dashed line marks the $(\Delta_0^{\text{FeSC}},\mu)=(0.1,1.54)$ case. FeSC-I-S junction: (e) and (f) $S(T,\Delta_0)$ and ZT$(T,\Delta_0)$, at $\mu=1.54$, $\Delta_0^{\text{FeSC}}=0.1$, and $\gamma=10^{-4}$. The dashed lines in panels (c) and (f) marks the values $\Delta^{th}_0=1.764T$. A cartoon in the top-right corner of each panel helps to recognise the type of junction considered at a glance.}
\label{FigureSM03}
\end{figure*}
For the sake of completeness, we also show how the Seebeck coefficient and TE efficiency behave considering a $d_{xy}$ pairing symmetry. As can be seen from Fig.~\ref{FigureSM03}, the main characteristics discussed in the manuscript for $s_\pm$-symmetry are also evident here, with a clear shift towards lower temperatures, in line with what was discussed in Fig.~3 in the main text for a nodal case. Another quite evident difference is that the sign of $S$ is reversed from what is shown in Fig.~2 of the main article, see Figs.~\ref{FigureSM03}(a)-(c)-(e): this indicates that where hole contributions for $s_\pm$-symmetry is predominant, particle contributions now prevail (and vice versa). In the case of an FeSC-I-N junction, we also note that the threshold doping value around which the sign of $S$ changes is somewhat higher than in the case shown in the article (the switch occurs roughly at $\mu\sim1.25$), see Fig.~\ref{FigureSM03}(b). Finally, we observe that the maximum values obtained for $S$ and ZT in this case are practically the same as those obtained in the case of $s_\pm$ pairing symmetry. However, considering an FeSC-I-S junction, the region of the parameter space where these maxima are reached is different, being the ZT peak in Fig.~\ref{FigureSM03}(f) located at $(T,\Delta_0)_{max}\simeq(0.23,2.1)\times10^{-4}$; in other words, in this case the maximum TE efficiency would be attainable for a superconducting tunnel junction at $T\simeq40\;\text{mK}$ formed between a FeSC and a SC with $T_c\simeq210\;\text{mK}$.

\begin{figure*}[t!!]
\centering
\includegraphics[width=\columnwidth]{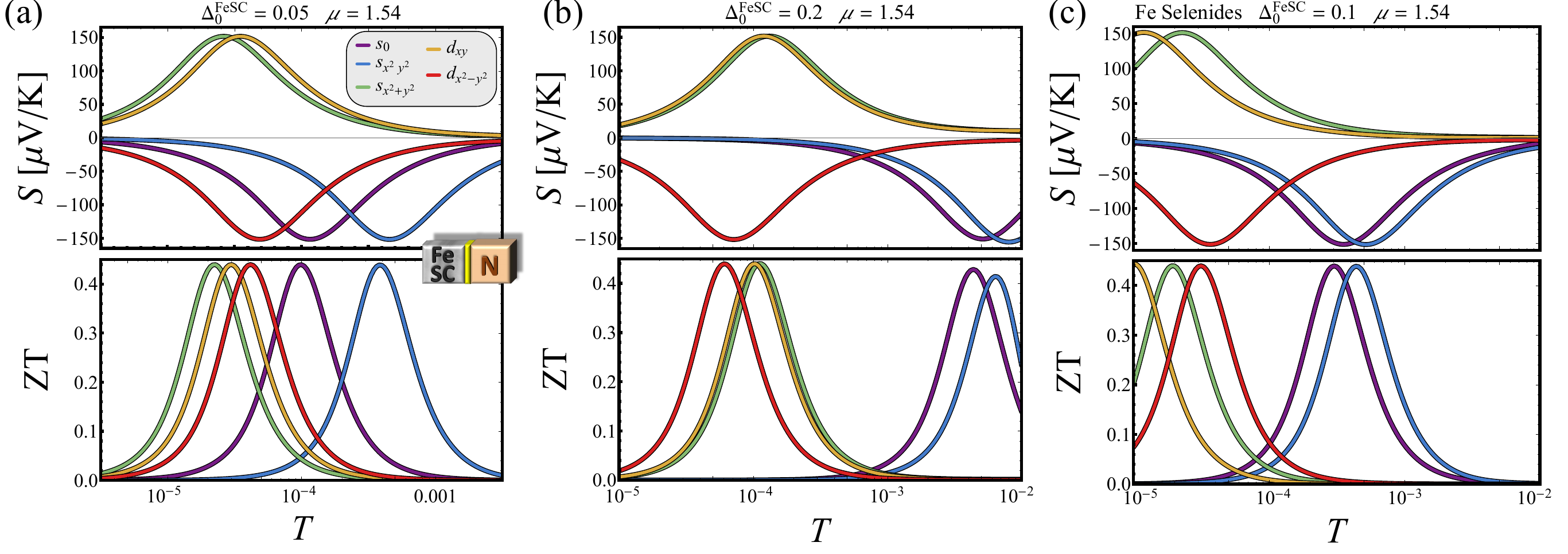}
\caption{FeSC-I-N junction: Temperature dependence of Seebeck coefficient, $S(T)$, (upper panels) and figure of merit, ZT$(T)$, (lower panels) by changing the symmetry of the order parameter, at a fixed $\mu=1.54$ and setting (a) $\Delta_0^{\text{FeSC}}=0.05$ and (b) $\Delta_0^{\text{FeSC}}=0.2$, while in (c) we set $\Delta_0^{\text{FeSC}}=0.1$ in the specific case of $(t_1,t_2,t_3,t_4) = (-1,1.5,-1.2-0.95)$ for iron selenides~\cite{Dum16}. The legend in panel (a) refers to all panels.}
\label{FigureSM05}
\end{figure*}

Finally, Fig.~\ref{FigureSM05} show that the TE of a tunnel junction resolves better between nodal and nodeless cases increasing the gap value, see panels (a) and (b) for $\Delta_0^{\text{FeSC}}=0.05$ and $0.2$, respectively. Furthermore, we tested the results also with the hopping parameters of FeSC selenides, see Fig.~\ref{FigureSM05}(c) for $\Delta_0^{\text{FeSC}}=0.1$ and $\mu=1.54$. 

%

%
\begin{figure*}[b!!]
\centering
\includegraphics[width=0.325\columnwidth]{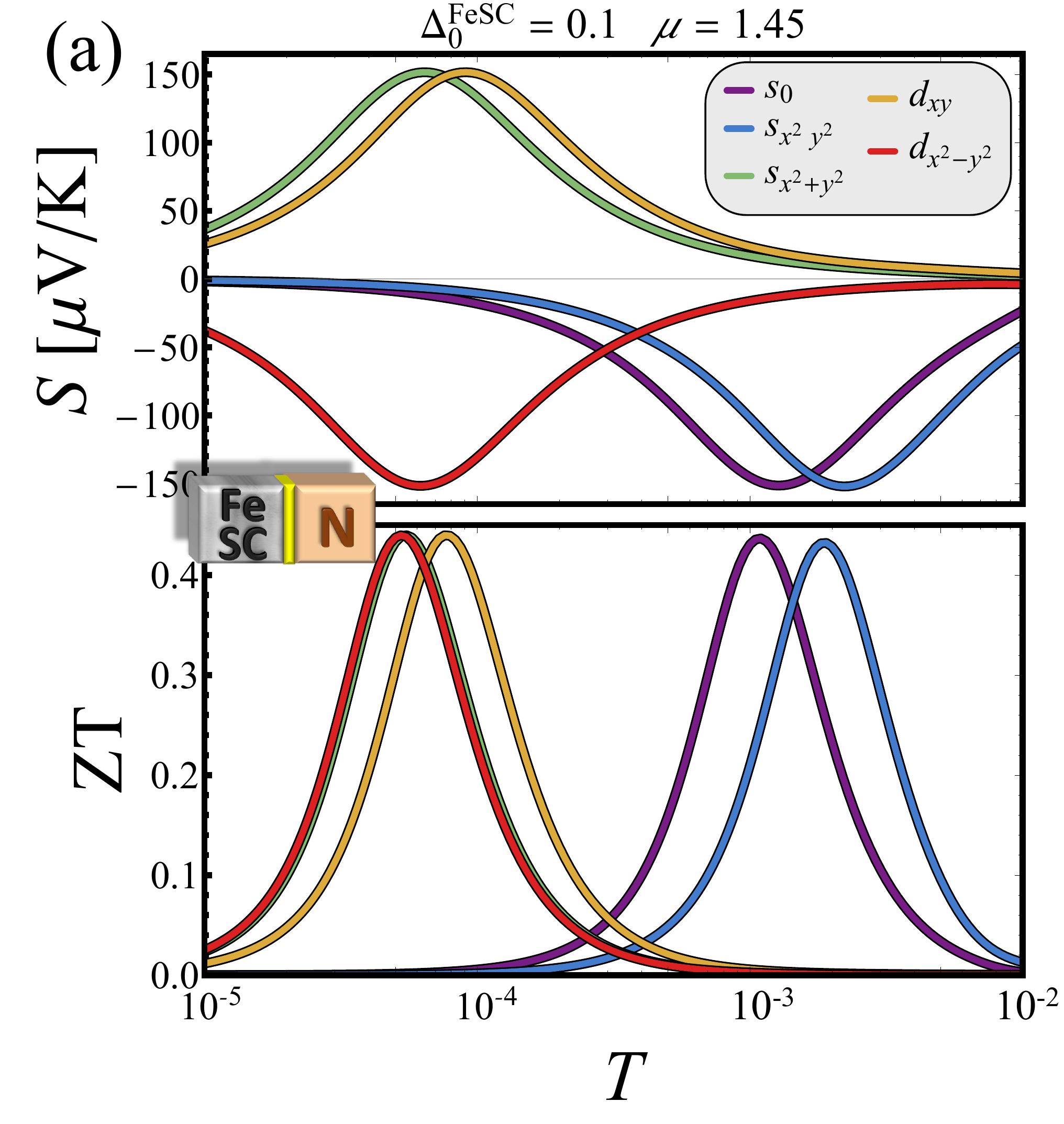}
\includegraphics[width=0.325\columnwidth]{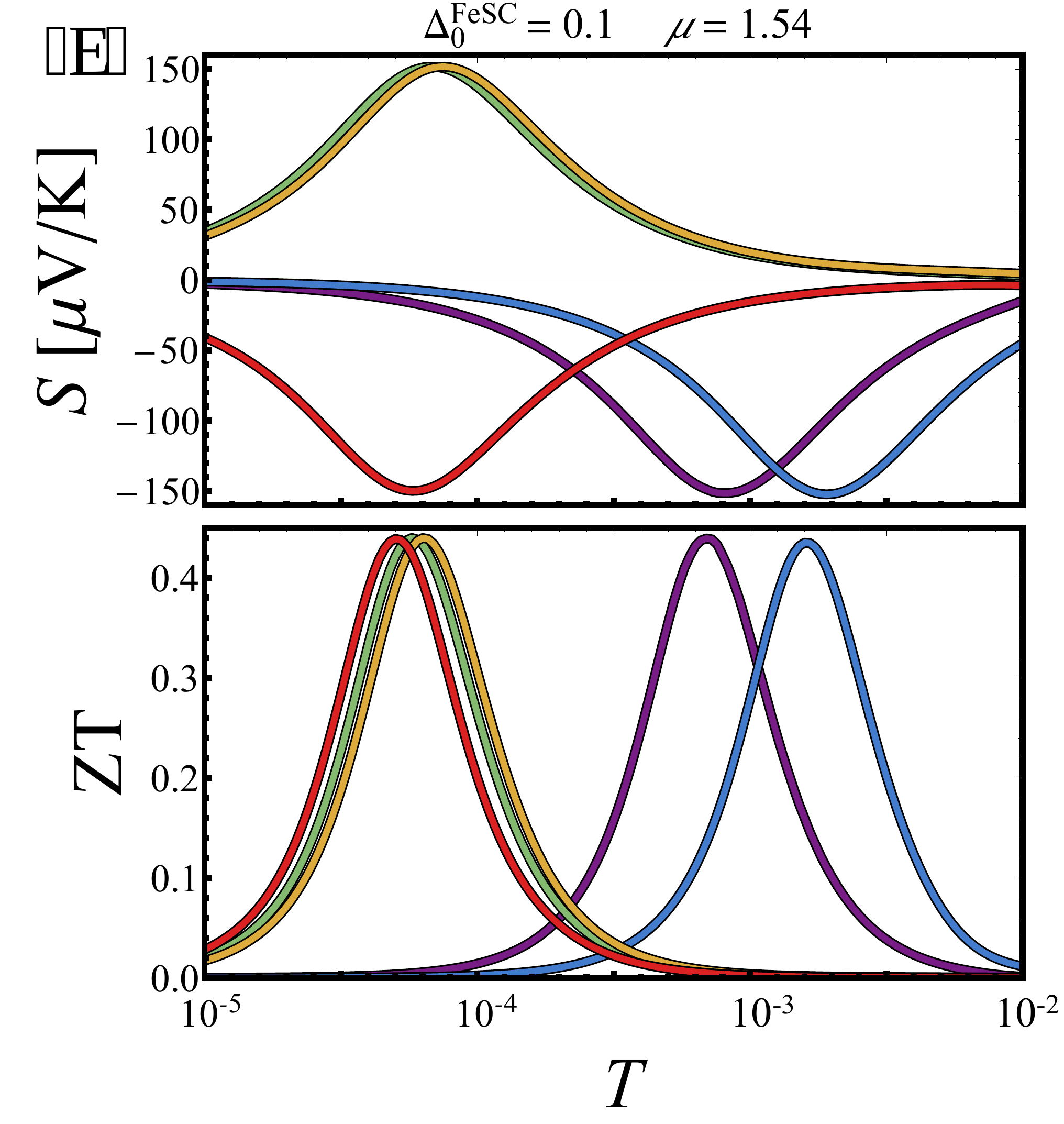}
\includegraphics[width=0.325\columnwidth]{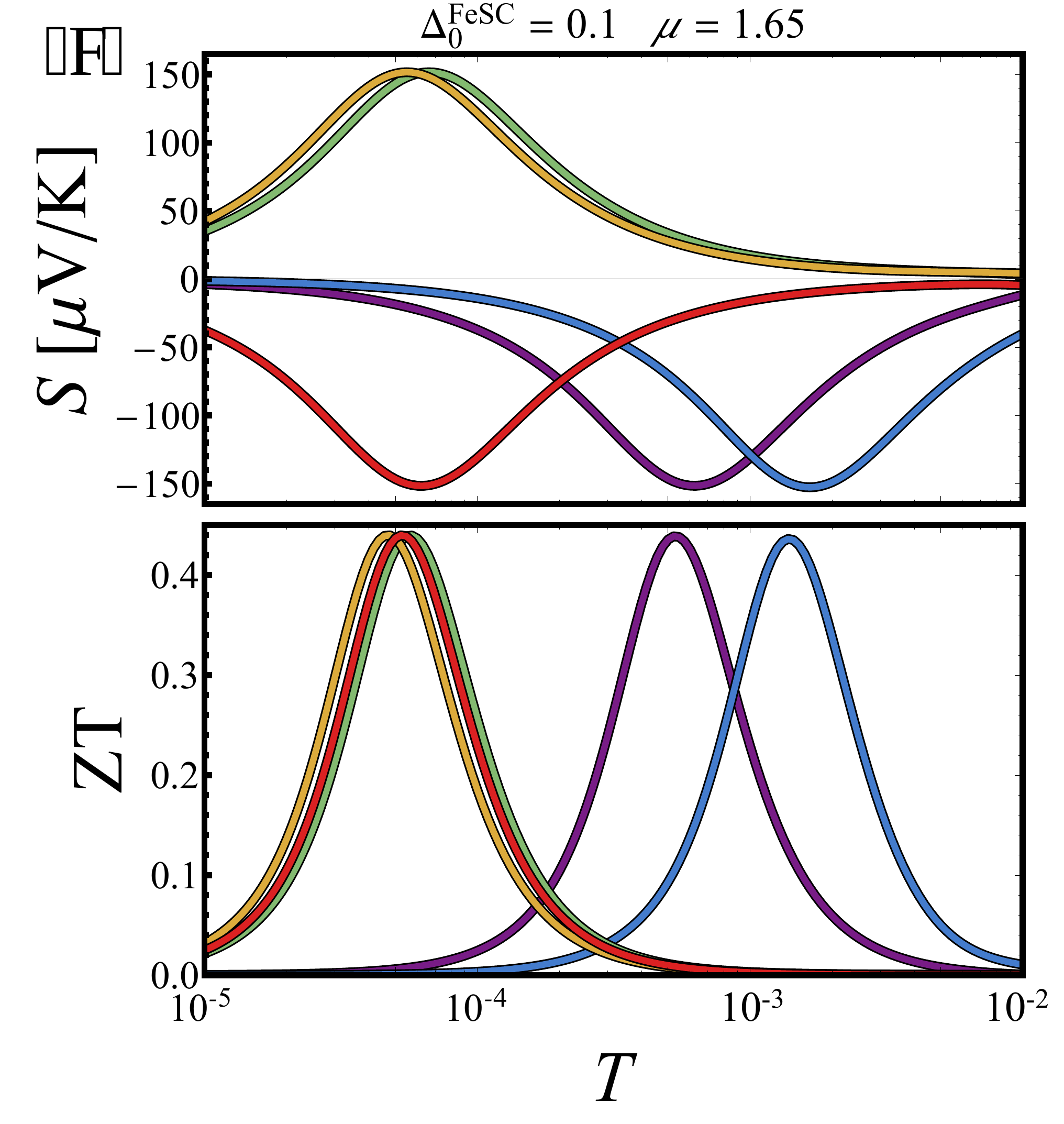}
\caption{FeSC-I-N junction: Temperature dependence of Seebeck coefficient, $S(T)$, (upper panels) and figure of merit, ZT$(T)$, (lower panels) by changing the symmetry of the order parameter, at a fixed $\Delta_0^{\text{FeSC}}=0.1$ and setting (a) $\mu=1.45$ and (b) $\mu=1.54$, while in (c) we set $\mu=1.65$. The legend in panel (a) refers to all panels.}
\label{FigureSM06}
\end{figure*}

\subsection{Robustness to chemical potential variations}

The level of doping usually considered in Fe-based superconductors, like FeSeTe, doesn't change dramatically the position of the chemical potential and we expect that the impact of $\mu$ on TE performance is not strong at changing the doping level by a few percent. We have verified that for a level doping of $3\%$ compared to the half-filling the chemical potential is roughly shifted by $1.45$ to $1.65$ and this does not affect the results. 
In any case, just to get an idea of how Fig.~3 of the main text changes in the case of a doped system, we show below the curves of $S(T)$ and ZT$(T)$ at different order parameter symmetries, for $\mu = \{1.45, 1.54, 1.65\}$. It is evident that the result is qualitatively robust to possible doping, although quantitatively there may be some differences.

\subsection{Power factor}

A complete TE characterization of the device requires also looking at the power that the system produces as an energy harvester. This can be quantified analyzing the electrical conductance $\sigma$, or more precisely the power factor $\text{PF}=\sigma\,S^2$, where $S$ is the Seebeck coefficient~\cite{Ben17}. Indeed, while large values of ZT imply a high efficiency, this often happens at the cost of low power, or alternatively if \text{PF} is very low, it may be experimentally difficult to measure $S$, because the voltage bias produced by the temperature gradient may drastically drop when a small load is applied to the system to measure such a voltage. However, defining the quantities of interest for our work, we focused on the intensive TE figures of merit: in fact, both the Seebeck coefficient and the ZT do not depend on the conductance of the tunnel barrier, $G_T$. The case of the power factor, which depends on $G_T$ being directly proportional to $\sigma$, is different.

\begin{figure*}[t!!]
\centering
\includegraphics[width=0.9\columnwidth]{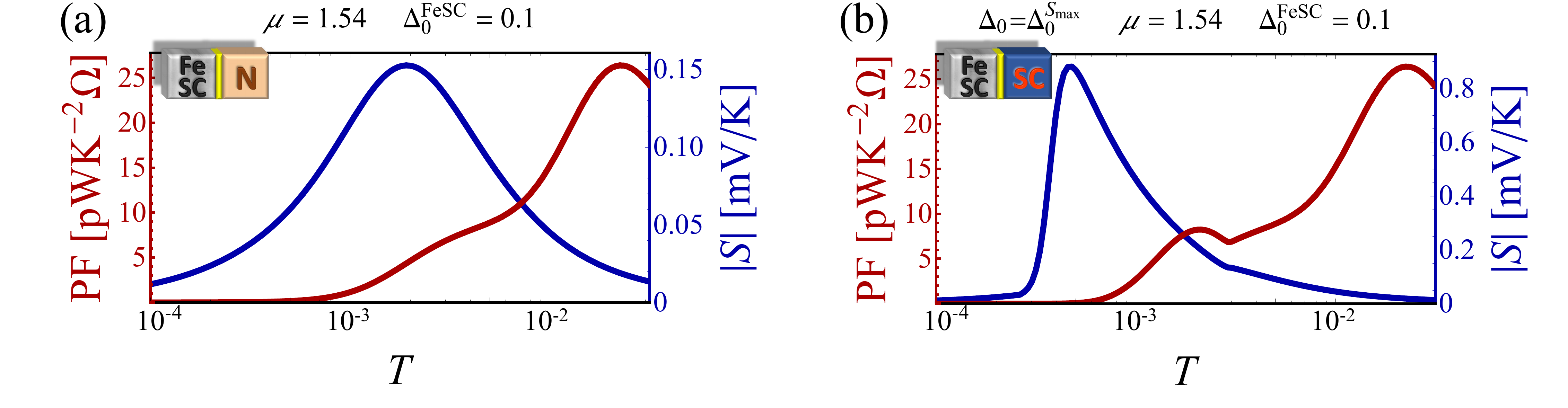}
\caption{Power factor (normalised to the tunnel conductance $G_T$ of the junction), $\text{PF}$, and Seebeck coefficient, $|S(T)|$ , as a function of temperature for an FeSC-I-N (a) and an FeSC-I-S junction (b). The values of other parameters are $\mu=1.54$ and $\Delta_0^{\text{FeSC}}=0.1$, while for (b) the value of $\Delta_0$ is chosen in such a way as to give the maximum of $S$. A cartoon in the top-right corner of each panel helps to recognize the type of junction considered at a glance.}
\label{FigureSM07}
\end{figure*}

In Fig.~\ref{FigureSM07} we report the comparison between $ \left| S \right|(T)$ and $\text{PF}(T)$, the latter normalised to the tunnel conductance of the junction. In particular, we focus on the cases of interest for our work, i.e., an FeSC-I-N junction, see panel (a), and an FeSC-I-S junction, see panel (b), in the case of an undoped FeSC, $\mu=1.54$, with $\Delta_0^{\text{FeSC}}=0.1$, i.e., the red dashed lines in Fig.~2(a-d) of the main text, while the value of $\Delta_0$ in (b) is chosen in such a way as to give the maximum of $\left| S \right|$.
We immediately notice that $\text{PF}(T)$, despite the quantitative differences, is qualitatively the same in the two cases, and that the power factor could actually be quite low at the maximum Seebeck coefficient. However, for our purpose the crucial point is if the effect is still measurable.
For an FeSC-I-N junction, i.e., in the cases described in Fig.~3 of the main text by comparing the various symmetries, it is evident that at the temperature where $ \left| S \right|$ peaks, the power factor is $\text{PF} \sim 5\; \text{pW}\, \text{K}^{-2}\,\Omega$. The case of the FeSC-I-S junction is different: in fact, the peak of $ \left| S \right|$ occurs at lower temperatures, where the power factor is quite low ($\text{PF} \sim 0.2\; \text{pW}\, \text{K}^{-2}\,\Omega$). In the latter case, a compromise between a not-too-low working temperature and a non-zero power factor may be necessary to measure ZT. We will return to the calculation of \text{PF} also in the next section.

\subsection{Relation between the superconducting gaps at the optimal TE efficiency}

\begin{figure*}[t!!]
\centering
\includegraphics[width=\columnwidth]{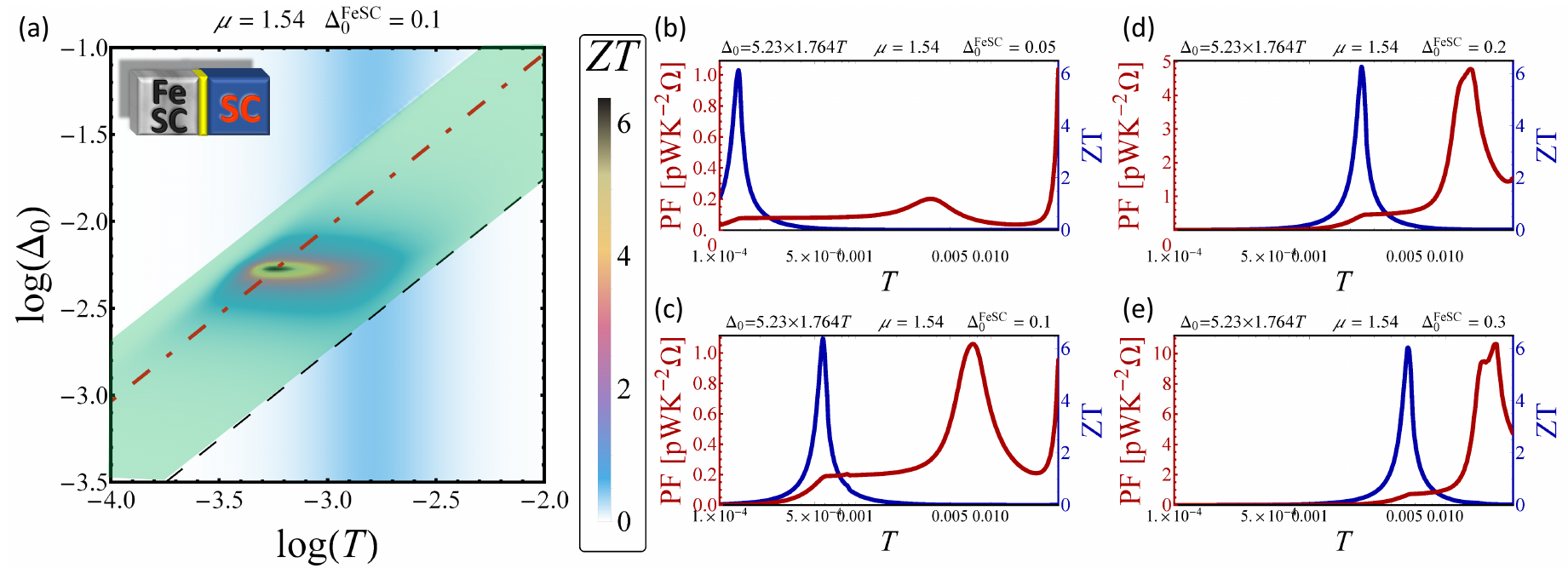}
\caption{FeSC-I-S junction: (a) ZT$(T,\Delta_0)$, at $\mu=1.54$ and $\Delta_0^{\text{FeSC}}=0.1$. The black dashed line mark the values $\Delta^{th}_0=1.764T$, while the red dot-dashed line is used to find the optimal TE point at different $\Delta_0^{\text{FeSC}}$ values, still within the green shaded area. A cartoon in the top-left corner helps to recognize the type of junction considered at a glance. (b-e) Temperature dependence of both the power factor, $\text{PF}$, (normalised to the tunnel conductance $G_T$ of the junction) and the ZT at the $\Delta_0$ values along the red dot-dashed line in (a), $\mu=1.54$, and $\Delta_0^{\text{FeSC}}= \{0.05, 0.1, 0.2, 0.3\}$. }
\label{FigureSM08}
\end{figure*}
\begin{figure*}[b!!]
\centering
\includegraphics[width=0.9\columnwidth]{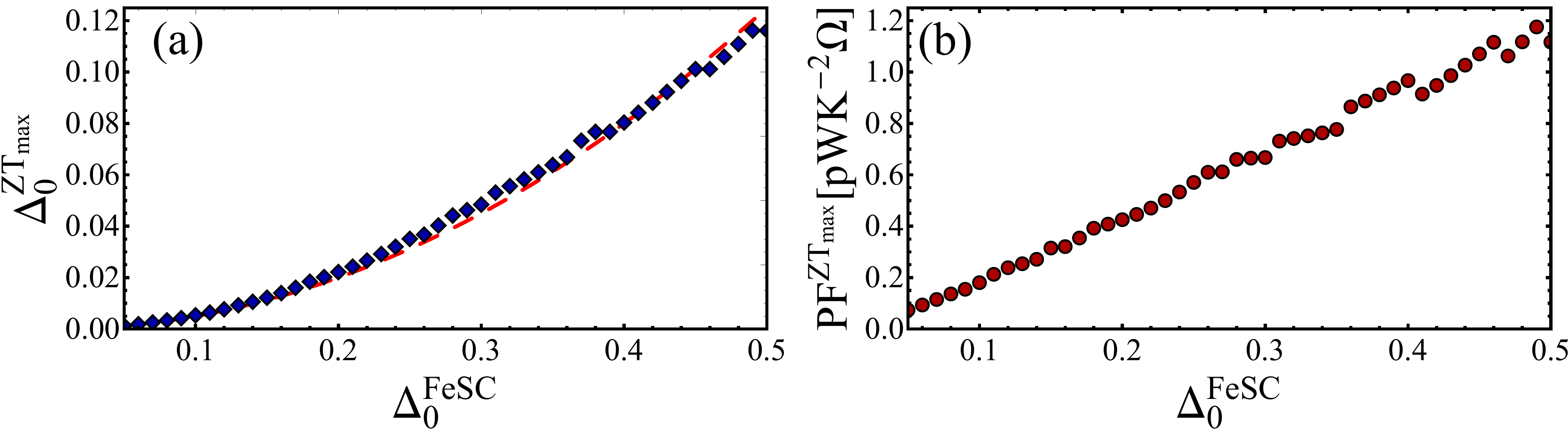}
\caption{FeSC-I-S junction: (a) $\Delta_0$ at the ZT maximum, $\Delta_0^{\text{ZT}_\text{max}}$, and (b) $\text{PF}$ (normalized to $G_T$) at the ZT maximum, $\text{PF}^{\text{ZT}_\text{max}}$, as a function of $\Delta_0^{\text{FeSC}}$. The red dashed curve in (a) represents the function $\Delta_0^{\text{ZT}_\text{max}} = k (\Delta_0^{\text{FeSC}})^2$, with $k\sim1/2$, that appears to fit our results very well.}
\label{FigureSM09}
\end{figure*}

One could in principle expect some correlation between the size of the gaps and the position of the peaks of the TE figures of merit, although the relation between these quantities is hard to guess given the nonlinear nature of the expressions involved. However, we realised that the relation connecting the superconducting gaps at the optimal TE point is not simply linear, but is instead a ``functional''; in particular, we found, at the maximum ZT, a quadratic dependence of $\Delta_0$ on $\Delta_0^{\text{FeSC}}$. The procedure that we follow to achieve this result is briefly outlined below.

First, we started looking at Fig.~2(f) of the main text, which is reported even in the panel (a) of Fig.~\ref{FigureSM08}, that is obtained for $\Delta_0^{\text{FeSC}}=01$. We argued that at a different $\Delta_0^{\text{FeSC}}$ the ``island'' of high ZT values would still emerge, but in another position within the highlighted green shaded area of the $(T,\Delta_0)$-parameter space, which is bounded by the black dotted line. The latter is a threshold for TE effects, i.e., it is the temperature above which the gap of the conventional superconductor is fully closed; thus, it does not dependent of the Fe-SC parameters. It is therefore reasonable to expect that the region of the $(T,\Delta_0)$-parameter space in which to expect the maximum TE effects is just ``before'' this threshold. We looked for the optimal TE conditions by ``staying'' along the red dot-dashed line, which is just parallel to the black dashed line and passing through the ZT maximum. In this way, for each $\Delta_0^{\text{FeSC}}$ it is possible to find the ``position'' of the ZT peak without scanning the whole $(T,\Delta_0)$-parameter space, thus avoiding a highly time-consuming numerical task. Panels (b-e) of Fig.~\ref{FigureSM08} are obtained in this way, setting four distinct values of $\Delta_0^{\text{FeSC}}= \{0.05, 0.1, 0.2, 0.3\}$; in each panel, we present both ZT$(T)$ (blue curve) and $\text{PF}(T)$ (red curve).
All these plots show a ZT$(T)$ peaked profile, reaching values $\gtrsim6$. At a given $\Delta_0^{\text{FeSC}}$, it is therefore easy to extract the position of the ZT peak (and therefore the $\Delta_0^{\text{ZT}_\text{max}}$ value) that gives the maximum TE efficiency. All that remains is to collect the pairs $(\Delta_0^{\text{FeSC}}, \Delta_0^{\text{ZT}_\text{max}})$ at different $\Delta_0^{\text{FeSC}}$ values.

In Fig.~\ref{FigureSM09} we report in (a) the $\Delta_0$ at the ZT maximum, $\Delta_0^{\text{ZT}_\text{max}}$, and in (b) the $\text{PF}$ at the ZT maximum, $\text{PF}^{\text{ZT}_\text{max}}$, (normalized to $G_T$) at different values of $\Delta_0^{\text{FeSC}}$. The red dashed curve in (a) describes the function $\Delta_0^{\text{ZT}_\text{max}} = k (\Delta_0^{\text{FeSC}})^2$, with $k\sim1/2$, that appears to fit our results very well. In (b), it is evident that $\text{PF}^{\text{ZT}_\text{max}}$ grows linearly with $\Delta_0^{\text{FeSC}}$.

In conclusion, we derived a quadratic dependence of the superconducting gap that gives high TE performance on the Fe-based superconducting gap. The reason for this dependence is not immediately apparent, and the question could be investigated further in more detail at a later stage in a following work.


\begin{thebibliography}{56}%
\makeatletter
\providecommand \@ifxundefined [1]{%
 \@ifx{#1\undefined}
}%
\providecommand \@ifnum [1]{%
 \ifnum #1\expandafter \@firstoftwo
 \else \expandafter \@secondoftwo
 \fi
}%
\providecommand \@ifx [1]{%
 \ifx #1\expandafter \@firstoftwo
 \else \expandafter \@secondoftwo
 \fi
}%
\providecommand \natexlab [1]{#1}%
\providecommand \enquote  [1]{``#1''}%
\providecommand \bibnamefont  [1]{#1}%
\providecommand \bibfnamefont [1]{#1}%
\providecommand \citenamefont [1]{#1}%
\providecommand \href@noop [0]{\@secondoftwo}%
\providecommand \href [0]{\begingroup \@sanitize@url \@href}%
\providecommand \@href[1]{\@@startlink{#1}\@@href}%
\providecommand \@@href[1]{\endgroup#1\@@endlink}%
\providecommand \@sanitize@url [0]{\catcode `\\12\catcode `\$12\catcode
  `\&12\catcode `\#12\catcode `\^12\catcode `\_12\catcode `\%12\relax}%
\providecommand \@@startlink[1]{}%
\providecommand \@@endlink[0]{}%
\providecommand \url  [0]{\begingroup\@sanitize@url \@url }%
\providecommand \@url [1]{\endgroup\@href {#1}{\urlprefix }}%
\providecommand \urlprefix  [0]{URL }%
\providecommand \Eprint [0]{\href }%
\providecommand \doibase [0]{http://dx.doi.org/}%
\providecommand \selectlanguage [0]{\@gobble}%
\providecommand \bibinfo  [0]{\@secondoftwo}%
\providecommand \bibfield  [0]{\@secondoftwo}%
\providecommand \translation [1]{[#1]}%
\providecommand \BibitemOpen [0]{}%
\providecommand \bibitemStop [0]{}%
\providecommand \bibitemNoStop [0]{.\EOS\space}%
\providecommand \EOS [0]{\spacefactor3000\relax}%
\providecommand \BibitemShut  [1]{\csname bibitem#1\endcsname}%
\let\auto@bib@innerbib\@empty
\bibitem [{\citenamefont {Fornieri}\ and\ \citenamefont
  {Giazotto}(2017)}]{For17}%
  \BibitemOpen
  \bibfield  {author} {\bibinfo {author} {\bibfnamefont {A.}~\bibnamefont
  {Fornieri}}\ and\ \bibinfo {author} {\bibfnamefont {F.}~\bibnamefont
  {Giazotto}},\ }\bibfield  {title} {\enquote {\bibinfo {title} {Towards
  phase-coherent caloritronics in superconducting circuits},}\ }\href {\doibase 10.1038/nnano.2017.204} {\bibfield  {journal}
  {\bibinfo  {journal} {Nat. Nanotechnol.}\ }\textbf {\bibinfo {volume} {12}},\
  \bibinfo {pages} {944} (\bibinfo {year} {2017})}\BibitemShut {NoStop}%
\bibitem [{\citenamefont {Hwang}\ and\ \citenamefont {Sothmann}(2020)}]{Hwa20}%
  \BibitemOpen
  \bibfield  {author} {\bibinfo {author} {\bibfnamefont {S.-Y.}\ \bibnamefont
  {Hwang}}\ and\ \bibinfo {author} {\bibfnamefont {B.}~\bibnamefont
  {Sothmann}},\ }\bibfield  {title} {\enquote {\bibinfo {title}
  {Phase-coherent caloritronics with ordinary and topological Josephson
  junctions},}\ }\href {\doibase 10.1140/epjst/e2019-900094-y} {\bibfield
  {journal} {\bibinfo  {journal} {Eur. Phys. J. Spec. Top.}\ }\textbf {\bibinfo
  {volume} {229}},\ \bibinfo {pages} {683} (\bibinfo {year}
  {2020})}\BibitemShut {NoStop}%
\bibitem [{\citenamefont {Giazotto}\ \emph {et~al.}(2006)\citenamefont
  {Giazotto}, \citenamefont {Heikkil\"a}, \citenamefont {Luukanen},
  \citenamefont {Savin},\ and\ \citenamefont {Pekola}}]{giazotto_rmp_2006}%
  \BibitemOpen
  \bibfield  {author} {\bibinfo {author} {\bibfnamefont {F.}~\bibnamefont
  {Giazotto}}, \bibinfo {author} {\bibfnamefont {T.~T.}\ \bibnamefont
  {Heikkil\"a}}, \bibinfo {author} {\bibfnamefont {A.}~\bibnamefont
  {Luukanen}}, \bibinfo {author} {\bibfnamefont {A.~M.}\ \bibnamefont {Savin}},
  \ and\ \bibinfo {author} {\bibfnamefont {J.~P.}\ \bibnamefont {Pekola}},\
  }\bibfield  {title} {\enquote {\bibinfo {title}
  {Opportunities for mesoscopics in thermometry and refrigeration: Physics and
  applications},}\ }\href {\doibase 10.1103/RevModPhys.78.217} {\bibfield  {journal} {\bibinfo
  {journal} {Rev. Mod. Phys.}\ }\textbf {\bibinfo {volume} {78}},\ \bibinfo
  {pages} {217} (\bibinfo {year} {2006})}\BibitemShut {NoStop}%
\bibitem [{\citenamefont {Muhonen}\ \emph {et~al.}(2012)\citenamefont
  {Muhonen}, \citenamefont {Meschke},\ and\ \citenamefont
  {Pekola}}]{pekola_2012}%
  \BibitemOpen
  \bibfield  {author} {\bibinfo {author} {\bibfnamefont {J.~T.}\ \bibnamefont
  {Muhonen}}, \bibinfo {author} {\bibfnamefont {M.}~\bibnamefont {Meschke}}, \
  and\ \bibinfo {author} {\bibfnamefont {J.~P.}\ \bibnamefont {Pekola}},\
  }\bibfield  {title} {\enquote {\bibinfo {title} {Micrometre-scale
  refrigerators},}\ }\href {\doibase 10.1088/0034-4885/75/4/046501} {\bibfield  {journal}
  {\bibinfo  {journal} {Rep. Prog. Phys.}\ }\textbf {\bibinfo {volume} {75}},\
  \bibinfo {pages} {046501} (\bibinfo {year} {2012})}\BibitemShut {NoStop}%
\bibitem [{\citenamefont {Ozaeta}\ \emph {et~al.}(2014)\citenamefont {Ozaeta},
  \citenamefont {Virtanen}, \citenamefont {Bergeret},\ and\ \citenamefont
  {Heikkil\"a}}]{Ozaeta2014}%
  \BibitemOpen
  \bibfield  {author} {\bibinfo {author} {\bibfnamefont {A.}~\bibnamefont
  {Ozaeta}}, \bibinfo {author} {\bibfnamefont {P.}~\bibnamefont {Virtanen}},
  \bibinfo {author} {\bibfnamefont {F.~S.}\ \bibnamefont {Bergeret}}, \ and\
  \bibinfo {author} {\bibfnamefont {T.~T.}\ \bibnamefont {Heikkil\"a}},\ }\bibfield  {title} {\enquote {\bibinfo {title} {Predicted very large
  thermoelectric effect in ferromagnet-superconductor junctions in the presence
  of a spin-splitting magnetic field},}\ }\href
  {\doibase 10.1103/PhysRevLett.112.057001} {\bibfield  {journal} {\bibinfo
  {journal} {Phys. Rev. Lett.}\ }\textbf {\bibinfo {volume} {112}},\ \bibinfo
  {pages} {057001} (\bibinfo {year} {2014})}\BibitemShut {NoStop}%
\bibitem [{\citenamefont {Kolenda}\ \emph
  {et~al.}(2016{\natexlab{a}})\citenamefont {Kolenda}, \citenamefont {Wolf},\
  and\ \citenamefont {Beckmann}}]{kolenda_2016}%
  \BibitemOpen
  \bibfield  {author} {\bibinfo {author} {\bibfnamefont {S.}~\bibnamefont
  {Kolenda}}, \bibinfo {author} {\bibfnamefont {M.~J.}\ \bibnamefont {Wolf}}, \
  and\ \bibinfo {author} {\bibfnamefont {D.}~\bibnamefont {Beckmann}},\ }\bibfield  {title} {\enquote {\bibinfo {title} {Observation of
  thermoelectric currents in high-field superconductor-ferromagnet tunnel
  junctions},}\ }\href
  {\doibase 10.1103/PhysRevLett.116.097001} {\bibfield  {journal} {\bibinfo
  {journal} {Phys. Rev. Lett.}\ }\textbf {\bibinfo {volume} {116}},\ \bibinfo
  {pages} {097001} (\bibinfo {year} {2016}{\natexlab{a}})}\BibitemShut
  {NoStop}%
\bibitem [{\citenamefont {Bergeret}\ \emph {et~al.}(2018)\citenamefont
  {Bergeret}, \citenamefont {Silaev}, \citenamefont {Virtanen},\ and\
  \citenamefont {Heikkil\"a}}]{Bergeret2018}%
  \BibitemOpen
  \bibfield  {author} {\bibinfo {author} {\bibfnamefont {F.~S.}\ \bibnamefont
  {Bergeret}}, \bibinfo {author} {\bibfnamefont {M.}~\bibnamefont {Silaev}},
  \bibinfo {author} {\bibfnamefont {P.}~\bibnamefont {Virtanen}}, \ and\
  \bibinfo {author} {\bibfnamefont {T.~T.}\ \bibnamefont {Heikkil\"a}},\ }\bibfield  {title} {\enquote {\bibinfo {title}
  {Colloquium: Nonequilibrium effects in superconductors with a spin-splitting
  field},}\ }\href
  {\doibase 10.1103/RevModPhys.90.041001} {\bibfield  {journal} {\bibinfo
  {journal} {Rev. Mod. Phys.}\ }\textbf {\bibinfo {volume} {90}},\ \bibinfo
  {pages} {041001} (\bibinfo {year} {2018})}\BibitemShut {NoStop}%
\bibitem [{\citenamefont {Hussein}\ \emph {et~al.}(2019)\citenamefont
  {Hussein}, \citenamefont {Governale}, \citenamefont {Kohler}, \citenamefont
  {Belzig}, \citenamefont {Giazotto},\ and\ \citenamefont
  {Braggio}}]{Hussein2019}%
  \BibitemOpen
  \bibfield  {author} {\bibinfo {author} {\bibfnamefont {R.}~\bibnamefont
  {Hussein}}, \bibinfo {author} {\bibfnamefont {M.}~\bibnamefont {Governale}},
  \bibinfo {author} {\bibfnamefont {S.}~\bibnamefont {Kohler}}, \bibinfo
  {author} {\bibfnamefont {W.}~\bibnamefont {Belzig}}, \bibinfo {author}
  {\bibfnamefont {F.}~\bibnamefont {Giazotto}}, \ and\ \bibinfo {author}
  {\bibfnamefont {A.}~\bibnamefont {Braggio}},\ }\bibfield  {title} {\enquote {\bibinfo {title}
  {Nonlocal thermoelectricity in a Cooper-pair splitter},}\ }\href {\doibase
  10.1103/PhysRevB.99.075429} {\bibfield  {journal} {\bibinfo  {journal} {Phys.
  Rev. B}\ }\textbf {\bibinfo {volume} {99}},\ \bibinfo {pages} {075429}
  (\bibinfo {year} {2019})}\BibitemShut {NoStop}%
\bibitem [{\citenamefont {Marchegiani}\ \emph {et~al.}(2020)\citenamefont
  {Marchegiani}, \citenamefont {Braggio},\ and\ \citenamefont
  {Giazotto}}]{Marchegiani2020}%
  \BibitemOpen
  \bibfield  {author} {\bibinfo {author} {\bibfnamefont {G.}~\bibnamefont
  {Marchegiani}}, \bibinfo {author} {\bibfnamefont {A.}~\bibnamefont
  {Braggio}}, \ and\ \bibinfo {author} {\bibfnamefont {F.}~\bibnamefont
  {Giazotto}},\ }\bibfield  {title} {\enquote {\bibinfo {title} {Nonlinear
  thermoelectricity with electron-hole symmetric systems},}\ }\href {\doibase 10.1103/PhysRevLett.124.106801} {\bibfield
  {journal} {\bibinfo  {journal} {Phys. Rev. Lett.}\ }\textbf {\bibinfo
  {volume} {124}},\ \bibinfo {pages} {106801} (\bibinfo {year}
  {2020})}\BibitemShut {NoStop}%
\bibitem [{\citenamefont {Blasi}\ \emph
  {et~al.}(2020{\natexlab{a}})\citenamefont {Blasi}, \citenamefont {Taddei},
  \citenamefont {Arrachea}, \citenamefont {Carrega},\ and\ \citenamefont
  {Braggio}}]{Blasi20}%
  \BibitemOpen
  \bibfield  {author} {\bibinfo {author} {\bibfnamefont {G.}~\bibnamefont
  {Blasi}}, \bibinfo {author} {\bibfnamefont {F.}~\bibnamefont {Taddei}},
  \bibinfo {author} {\bibfnamefont {L.}~\bibnamefont {Arrachea}}, \bibinfo
  {author} {\bibfnamefont {M.}~\bibnamefont {Carrega}}, \ and\ \bibinfo
  {author} {\bibfnamefont {A.}~\bibnamefont {Braggio}},\ }\bibfield  {title} {\enquote {\bibinfo {title}
  {Nonlocal thermoelectricity in a
  superconductor--topological-insulator--superconductor junction in contact
  with a normal-metal probe: Evidence for helical edge states},}\ }\href {\doibase
  10.1103/PhysRevLett.124.227701} {\bibfield  {journal} {\bibinfo  {journal}
  {Phys. Rev. Lett.}\ }\textbf {\bibinfo {volume} {124}},\ \bibinfo {pages}
  {227701} (\bibinfo {year} {2020}{\natexlab{a}})}\BibitemShut {NoStop}%
\bibitem [{\citenamefont {Blasi}\ \emph
  {et~al.}(2020{\natexlab{b}})\citenamefont {Blasi}, \citenamefont {Taddei},
  \citenamefont {Arrachea}, \citenamefont {Carrega},\ and\ \citenamefont
  {Braggio}}]{Blasi20a}%
  \BibitemOpen
  \bibfield  {author} {\bibinfo {author} {\bibfnamefont {G.}~\bibnamefont
  {Blasi}}, \bibinfo {author} {\bibfnamefont {F.}~\bibnamefont {Taddei}},
  \bibinfo {author} {\bibfnamefont {L.}~\bibnamefont {Arrachea}}, \bibinfo
  {author} {\bibfnamefont {M.}~\bibnamefont {Carrega}}, \ and\ \bibinfo
  {author} {\bibfnamefont {A.}~\bibnamefont {Braggio}},\ }\bibfield  {title} {\enquote {\bibinfo {title}
  {Nonlocal thermoelectricity in a topological Andreev interferometer},}\
  }\href {\doibase
  10.1103/PhysRevB.102.241302} {\bibfield  {journal} {\bibinfo  {journal}
  {Phys. Rev. B}\ }\textbf {\bibinfo {volume} {102}},\ \bibinfo {pages}
  {241302} (\bibinfo {year} {2020}{\natexlab{b}})}\BibitemShut {NoStop}%
\bibitem [{\citenamefont {Germanese}\ \emph {et~al.}(2022)\citenamefont
  {Germanese}, \citenamefont {Paolucci}, \citenamefont {Marchegiani},
  \citenamefont {Braggio},\ and\ \citenamefont {Giazotto}}]{Germanese2022}%
  \BibitemOpen
  \bibfield  {author} {\bibinfo {author} {\bibfnamefont {G.}~\bibnamefont
  {Germanese}}, \bibinfo {author} {\bibfnamefont {F.}~\bibnamefont {Paolucci}},
  \bibinfo {author} {\bibfnamefont {G.}~\bibnamefont {Marchegiani}}, \bibinfo
  {author} {\bibfnamefont {A.}~\bibnamefont {Braggio}}, \ and\ \bibinfo
  {author} {\bibfnamefont {F.}~\bibnamefont {Giazotto}},\ }\bibfield  {title} {\enquote {\bibinfo {title} {Bipolar
  thermoelectric Josephson engine},}\ }\href {\doibase
  10.1038/s41565-022-01208-y} {\bibfield  {journal} {\bibinfo  {journal} {Nat.
  Nanotechnol.}\ }\textbf {\bibinfo {volume} {17}},\ \bibinfo {pages} {1084}
  (\bibinfo {year} {2022})}\BibitemShut {NoStop}%
\bibitem [{\citenamefont {Giazotto}\ \emph {et~al.}(2015)\citenamefont
  {Giazotto}, \citenamefont {Solinas}, \citenamefont {Braggio},\ and\
  \citenamefont {Bergeret}}]{Giazotto2015}%
  \BibitemOpen
  \bibfield  {author} {\bibinfo {author} {\bibfnamefont {F.}~\bibnamefont
  {Giazotto}}, \bibinfo {author} {\bibfnamefont {P.}~\bibnamefont {Solinas}},
  \bibinfo {author} {\bibfnamefont {A.}~\bibnamefont {Braggio}}, \ and\
  \bibinfo {author} {\bibfnamefont {F.~S.}\ \bibnamefont {Bergeret}},\ }\bibfield  {title} {\enquote {\bibinfo {title}
  {Ferromagnetic-insulator-based superconducting junctions as sensitive
  electron thermometers},}\ }\href
  {\doibase 10.1103/PhysRevApplied.4.044016} {\bibfield  {journal} {\bibinfo
  {journal} {Phys. Rev. Appl.}\ }\textbf {\bibinfo {volume} {4}},\ \bibinfo
  {pages} {044016} (\bibinfo {year} {2015})}\BibitemShut {NoStop}%
\bibitem [{\citenamefont {Heikkil\"a}\ \emph {et~al.}(2018)\citenamefont
  {Heikkil\"a}, \citenamefont {Ojaj\"arvi}, \citenamefont {Maasilta},
  \citenamefont {Strambini}, \citenamefont {Giazotto},\ and\ \citenamefont
  {Bergeret}}]{Hekkila2018}%
  \BibitemOpen
  \bibfield  {author} {\bibinfo {author} {\bibfnamefont {T.~T.}\ \bibnamefont
  {Heikkil\"a}}, \bibinfo {author} {\bibfnamefont {R.}~\bibnamefont
  {Ojaj\"arvi}}, \bibinfo {author} {\bibfnamefont {I.~J.}\ \bibnamefont
  {Maasilta}}, \bibinfo {author} {\bibfnamefont {E.}~\bibnamefont {Strambini}},
  \bibinfo {author} {\bibfnamefont {F.}~\bibnamefont {Giazotto}}, \ and\
  \bibinfo {author} {\bibfnamefont {F.~S.}\ \bibnamefont {Bergeret}},\ }\bibfield  {title} {\enquote {\bibinfo {title} {Thermoelectric radiation
  detector based on superconductor-ferromagnet systems},}\ }\href
  {\doibase 10.1103/PhysRevApplied.10.034053} {\bibfield  {journal} {\bibinfo
  {journal} {Phys. Rev. Appl.}\ }\textbf {\bibinfo {volume} {10}},\ \bibinfo
  {pages} {034053} (\bibinfo {year} {2018})}\BibitemShut {NoStop}%
\bibitem [{\citenamefont {Paolucci}\ \emph {et~al.}(2023)\citenamefont
  {Paolucci}, \citenamefont {Germanese}, \citenamefont {Braggio},\ and\
  \citenamefont {Giazotto}}]{Paolucci2023}%
  \BibitemOpen
  \bibfield  {author} {\bibinfo {author} {\bibfnamefont {F.}~\bibnamefont
  {Paolucci}}, \bibinfo {author} {\bibfnamefont {G.}~\bibnamefont {Germanese}},
  \bibinfo {author} {\bibfnamefont {A.}~\bibnamefont {Braggio}}, \ and\
  \bibinfo {author} {\bibfnamefont {F.}~\bibnamefont {Giazotto}},\ }\bibfield  {title} {\enquote {\bibinfo {title} {A highly sensitive broadband superconducting thermoelectric single-photon detector},}\ }\href
  {\doibase 10.1063/5.0145544} {\bibfield  {journal} {\bibinfo  {journal}
  {Applied Physics Letters}\ }\textbf {\bibinfo {volume} {122}},\ \bibinfo
  {pages} {173503} (\bibinfo {year} {2023})}\BibitemShut {NoStop}%
\bibitem [{\citenamefont {Machon}\ \emph {et~al.}(2014)\citenamefont {Machon},
  \citenamefont {Eschrig},\ and\ \citenamefont {Belzig}}]{Machon2014}%
  \BibitemOpen
  \bibfield  {author} {\bibinfo {author} {\bibfnamefont {P.}~\bibnamefont
  {Machon}}, \bibinfo {author} {\bibfnamefont {M.}~\bibnamefont {Eschrig}}, \
  and\ \bibinfo {author} {\bibfnamefont {W.}~\bibnamefont {Belzig}},\
  }\bibfield  {title} {\enquote {\bibinfo {title} {Giant
  thermoelectric effects in a proximity-coupled superconductor--ferromagnet
  device},}\ }\href@noop {} {\bibfield  {journal} {\bibinfo  {journal} {New J. Phys.}\
  }\textbf {\bibinfo {volume} {16}},\ \bibinfo {pages} {073002} (\bibinfo
  {year} {2014})}\BibitemShut {NoStop}%
\bibitem [{\citenamefont {Goll}(2006)}]{Goll2006Book}%
  \BibitemOpen
  \bibfield  {author} {\bibinfo {author} {\bibfnamefont {G.}~\bibnamefont
  {Goll}},\ }\href {\doibase 10.1007/11010715} {\emph {\bibinfo {title}
  {Unconventional Superconductors: Experimental Investigation of the
  Order-Parameter Symmetry of Unconventional Superconductors}}}\ (\bibinfo
  {publisher} {Springer Berlin Heidelberg},\ \bibinfo {address} {Berlin,
  Heidelberg},\ \bibinfo {year} {2006})\BibitemShut {NoStop}%
  \bibitem [{\citenamefont {Reid}\ \emph {et~al.}(2012)\citenamefont {Reid},
  \citenamefont {Tanatar}, \citenamefont {Juneau-Fecteau}, \citenamefont
  {Gordon}, \citenamefont {de~Cotret}, \citenamefont {Doiron-Leyraud},
  \citenamefont {Saito}, \citenamefont {Fukazawa}, \citenamefont {Kohori},
  \citenamefont {Kihou}, \citenamefont {Lee}, \citenamefont {Iyo},
  \citenamefont {Eisaki}, \citenamefont {Prozorov},\ and\ \citenamefont
  {Taillefer}}]{Rei12}%
  \BibitemOpen
  \bibfield  {author} {\bibinfo {author} {\bibfnamefont {J.-P.}\ \bibnamefont
  {Reid}}, \bibinfo {author} {\bibfnamefont {M.~A.}\ \bibnamefont {Tanatar}},
  \bibinfo {author} {\bibfnamefont {A.}~\bibnamefont {Juneau-Fecteau}},
  \bibinfo {author} {\bibfnamefont {R.~T.}\ \bibnamefont {Gordon}}, \bibinfo
  {author} {\bibfnamefont {S.~R.}\ \bibnamefont {de~Cotret}}, \bibinfo {author}
  {\bibfnamefont {N.}~\bibnamefont {Doiron-Leyraud}}, \bibinfo {author}
  {\bibfnamefont {T.}~\bibnamefont {Saito}}, \bibinfo {author} {\bibfnamefont
  {H.}~\bibnamefont {Fukazawa}}, \bibinfo {author} {\bibfnamefont
  {Y.}~\bibnamefont {Kohori}}, \bibinfo {author} {\bibfnamefont
  {K.}~\bibnamefont {Kihou}}, \bibinfo {author} {\bibfnamefont {C.~H.}\
  \bibnamefont {Lee}}, \bibinfo {author} {\bibfnamefont {A.}~\bibnamefont
  {Iyo}}, \bibinfo {author} {\bibfnamefont {H.}~\bibnamefont {Eisaki}},
  \bibinfo {author} {\bibfnamefont {R.}~\bibnamefont {Prozorov}}, \ and\
  \bibinfo {author} {\bibfnamefont {L.}~\bibnamefont {Taillefer}},\ }\href
  {\doibase 10.1103/PhysRevLett.109.087001} {\bibfield  {journal} {\bibinfo
  {journal} {Phys. Rev. Lett.}\ }\textbf {\bibinfo {volume} {109}},\ \bibinfo
  {pages} {087001} (\bibinfo {year} {2012})}\BibitemShut {NoStop}%
\bibitem [{\citenamefont {Citro}\ and\ \citenamefont
  {Mancini}(2017)}]{Citro2017Book}%
  \BibitemOpen
  \bibfield  {author} {\bibinfo {author} {\bibfnamefont {R.}~\bibnamefont
  {Citro}}\ and\ \bibinfo {author} {\bibfnamefont {F.}~\bibnamefont
  {Mancini}},\ }\href {\doibase 10.1007/978-3-319-56117-2} {\emph {\bibinfo
  {title} {The Iron Pnictide Superconductors: An Introduction and Overview}}},\
  edited by\ \bibinfo {editor} {\bibfnamefont {F.}~\bibnamefont {Mancini}}\
  and\ \bibinfo {editor} {\bibfnamefont {R.}~\bibnamefont {Citro}}\ (\bibinfo
  {publisher} {Springer International Publishing},\ \bibinfo {address} {Cham},\
  \bibinfo {year} {2017})\BibitemShut {NoStop}%
  \bibitem [{\citenamefont {Benjamin}\ and\ \citenamefont
  {Mohapatra}(2021)}]{Ben20}%
  \BibitemOpen
  \bibfield  {author} {\bibinfo {author} {\bibfnamefont {C.}~\bibnamefont
  {Benjamin}}\ and\ \bibinfo {author} {\bibfnamefont {T.}~\bibnamefont
  {Mohapatra}},\ }\href {\doibase 10.1209/0295-5075/132/47002} {\bibfield
  {journal} {\bibinfo  {journal} {Europhysics Letters}\ }\textbf {\bibinfo
  {volume} {132}},\ \bibinfo {pages} {47002} (\bibinfo {year}
  {2021})}\BibitemShut {NoStop}%
\bibitem [{\citenamefont {Mazin}\ \emph {et~al.}(2008)\citenamefont {Mazin},
  \citenamefont {Singh}, \citenamefont {Johannes},\ and\ \citenamefont
  {Du}}]{mazin_2008}%
  \BibitemOpen
  \bibfield  {author} {\bibinfo {author} {\bibfnamefont {I.~I.}\ \bibnamefont
  {Mazin}}, \bibinfo {author} {\bibfnamefont {D.~J.}\ \bibnamefont {Singh}},
  \bibinfo {author} {\bibfnamefont {M.~D.}\ \bibnamefont {Johannes}}, \ and\
  \bibinfo {author} {\bibfnamefont {M.~H.}\ \bibnamefont {Du}},\ }\bibfield
  {title} {\enquote {\bibinfo {title} {Unconventional superconductivity with a
  sign reversal in the order parameter of
  LaFeAsO$_{1-x}$F$_{x}$},}\ }\href
  {\doibase 10.1103/PhysRevLett.101.057003} {\bibfield  {journal} {\bibinfo
  {journal} {Phys. Rev. Lett.}\ }\textbf {\bibinfo {volume} {101}},\ \bibinfo
  {pages} {057003} (\bibinfo {year} {2008})}\BibitemShut {NoStop}%
\bibitem [{\citenamefont {Wang}\ \emph {et~al.}(2009)\citenamefont {Wang},
  \citenamefont {Zhai}, \citenamefont {Ran}, \citenamefont {Vishwanath},\ and\
  \citenamefont {Lee}}]{wang_2009}%
  \BibitemOpen
  \bibfield  {author} {\bibinfo {author} {\bibfnamefont {F.}~\bibnamefont
  {Wang}}, \bibinfo {author} {\bibfnamefont {H.}~\bibnamefont {Zhai}}, \bibinfo
  {author} {\bibfnamefont {Y.}~\bibnamefont {Ran}}, \bibinfo {author}
  {\bibfnamefont {A.}~\bibnamefont {Vishwanath}}, \ and\ \bibinfo {author}
  {\bibfnamefont {D.-H.}\ \bibnamefont {Lee}},\ }\bibfield  {title}
  {\enquote {\bibinfo {title} {Functional renormalization-group study of the
  pairing symmetry and pairing mechanism of the FeAs-based high-temperature
  superconductor},}\ }\href {\doibase
  10.1103/PhysRevLett.102.047005} {\bibfield  {journal} {\bibinfo  {journal}
  {Phys. Rev. Lett.}\ }\textbf {\bibinfo {volume} {102}},\ \bibinfo {pages}
  {047005} (\bibinfo {year} {2009})}\BibitemShut {NoStop}%
\bibitem [{\citenamefont {Guarcello}\ and\ \citenamefont
  {Citro}(2021)}]{Gua21}%
  \BibitemOpen
  \bibfield  {author} {\bibinfo {author} {\bibfnamefont {C.}~\bibnamefont
  {Guarcello}}\ and\ \bibinfo {author} {\bibfnamefont {R.}~\bibnamefont
  {Citro}},\ }\bibfield  {title} {\enquote {\bibinfo {title} {Progresses on
  topological phenomena, time-driven phase transitions, and unconventional
  superconductivity},}\ }\href {\doibase 10.1209/0295-5075/132/60003} {\bibfield
  {journal} {\bibinfo  {journal} {Europhys. Lett.}\ }\textbf {\bibinfo {volume}
  {132}},\ \bibinfo {pages} {60003} (\bibinfo {year} {2021})}\BibitemShut
  {NoStop}%
\bibitem [{\citenamefont {Kuroki}\ \emph {et~al.}(2008)\citenamefont {Kuroki},
  \citenamefont {Onari}, \citenamefont {Arita}, \citenamefont {Usui},
  \citenamefont {Tanaka}, \citenamefont {Kontani},\ and\ \citenamefont
  {Aoki}}]{kontani_2008}%
  \BibitemOpen
  \bibfield  {author} {\bibinfo {author} {\bibfnamefont {K.}~\bibnamefont
  {Kuroki}}, \bibinfo {author} {\bibfnamefont {S.}~\bibnamefont {Onari}},
  \bibinfo {author} {\bibfnamefont {R.}~\bibnamefont {Arita}}, \bibinfo
  {author} {\bibfnamefont {H.}~\bibnamefont {Usui}}, \bibinfo {author}
  {\bibfnamefont {Y.}~\bibnamefont {Tanaka}}, \bibinfo {author} {\bibfnamefont
  {H.}~\bibnamefont {Kontani}}, \ and\ \bibinfo {author} {\bibfnamefont
  {H.}~\bibnamefont {Aoki}},\ }\bibfield  {title} {\enquote {\bibinfo {title} {Unconventional
  pairing originating from the disconnected Fermi surfaces of superconducting
  LaFeAsO$_{1-x}$F$_{x}$},}\ } \href {\doibase 10.1103/PhysRevLett.101.087004}
  {\bibfield  {journal} {\bibinfo  {journal} {Phys. Rev. Lett.}\ }\textbf
  {\bibinfo {volume} {101}},\ \bibinfo {pages} {087004} (\bibinfo {year}
  {2008})}\BibitemShut {NoStop}%
\bibitem [{\citenamefont {Chen}\ \emph {et~al.}(2008)\citenamefont {Chen},
  \citenamefont {Tesanovic}, \citenamefont {Liu}, \citenamefont {Chen},\ and\
  \citenamefont {Chien}}]{chen_2008}%
  \BibitemOpen
  \bibfield  {author} {\bibinfo {author} {\bibfnamefont {T.}~\bibnamefont
  {Chen}}, \bibinfo {author} {\bibfnamefont {Z.}~\bibnamefont {Tesanovic}},
  \bibinfo {author} {\bibfnamefont {R.}~\bibnamefont {Liu}}, \bibinfo {author}
  {\bibfnamefont {X.}~\bibnamefont {Chen}}, \ and\ \bibinfo {author}
  {\bibfnamefont {C.}~\bibnamefont {Chien}},\ }\bibfield  {title} {\enquote
  {\bibinfo {title} {A BCS-like gap in the superconductor SmFeAsO$_{0. 85}$F$_{0.15}$},}\ }\href@noop {} {\bibfield
  {journal} {\bibinfo  {journal} {Nature}\ }\textbf {\bibinfo {volume} {453}},\
  \bibinfo {pages} {1224} (\bibinfo {year} {2008})}\BibitemShut {NoStop}%
\bibitem [{\citenamefont {Daghero}\ \emph {et~al.}(2012)\citenamefont
  {Daghero}, \citenamefont {Tortello}, \citenamefont {Ummarino}, \citenamefont
  {Stepanov}, \citenamefont {Bernardini}, \citenamefont {Tropeano},
  \citenamefont {Putti},\ and\ \citenamefont {Gonnelli}}]{daghero_2012}%
  \BibitemOpen
  \bibfield  {author} {\bibinfo {author} {\bibfnamefont {D.}~\bibnamefont
  {Daghero}}, \bibinfo {author} {\bibfnamefont {M.}~\bibnamefont {Tortello}},
  \bibinfo {author} {\bibfnamefont {G.}~\bibnamefont {Ummarino}}, \bibinfo
  {author} {\bibfnamefont {V.}~\bibnamefont {Stepanov}}, \bibinfo {author}
  {\bibfnamefont {F.}~\bibnamefont {Bernardini}}, \bibinfo {author}
  {\bibfnamefont {M.}~\bibnamefont {Tropeano}}, \bibinfo {author}
  {\bibfnamefont {M.}~\bibnamefont {Putti}}, \ and\ \bibinfo {author}
  {\bibfnamefont {R.}~\bibnamefont {Gonnelli}},\ }\bibfield  {title} {\enquote {\bibinfo
  {title} {Effects of isoelectronic Ru substitution at the Fe site on the
  energy gaps of optimally f-doped SmFeAsO},}\ }\href@noop {} {\bibfield
  {journal} {\bibinfo  {journal} {Supercond. Sci. Technol.}\ }\textbf {\bibinfo
  {volume} {25}},\ \bibinfo {pages} {084012} (\bibinfo {year}
  {2012})}\BibitemShut {NoStop}%
\bibitem [{\citenamefont {Pallecchi}\ \emph {et~al.}(2016)\citenamefont
  {Pallecchi}, \citenamefont {Caglieris},\ and\ \citenamefont
  {Putti}}]{Pallecchi2016}%
  \BibitemOpen
  \bibfield  {author} {\bibinfo {author} {\bibfnamefont {I.}~\bibnamefont
  {Pallecchi}}, \bibinfo {author} {\bibfnamefont {F.}~\bibnamefont
  {Caglieris}}, \ and\ \bibinfo {author} {\bibfnamefont {M.}~\bibnamefont
  {Putti}},\ }\bibfield  {title} {\enquote {\bibinfo {title} {Thermoelectric
  properties of iron-based superconductors and parent compounds},}\ }\href@noop {} {\bibfield  {journal} {\bibinfo  {journal}
  {Supercond. Sci. Technol.}\ }\textbf {\bibinfo {volume} {29}},\ \bibinfo
  {pages} {073002} (\bibinfo {year} {2016})}\BibitemShut {NoStop}%
\bibitem [{\citenamefont {Ziman}(2001)}]{Ziman2001}%
  \BibitemOpen
  \bibfield  {author} {\bibinfo {author} {\bibfnamefont {J.~M.}\ \bibnamefont
  {Ziman}},\ }\href@noop {} {\emph {\bibinfo {title} {Electrons and phonons:
  the theory of transport phenomena in solids}}}\ (\bibinfo  {publisher}
  {Oxford university press},\ \bibinfo {year} {2001})\BibitemShut {NoStop}%
\bibitem [{\citenamefont {Caglieris}\ \emph {et~al.}(2014)\citenamefont
  {Caglieris}, \citenamefont {Braggio}, \citenamefont {Pallecchi},
  \citenamefont {Provino}, \citenamefont {Pani}, \citenamefont {Lamura},
  \citenamefont {Jost}, \citenamefont {Zeitler}, \citenamefont
  {Galleani~D'Agliano}, \citenamefont {Manfrinetti},\ and\ \citenamefont
  {Putti}}]{Caglieris2014}%
  \BibitemOpen
  \bibfield  {author} {\bibinfo {author} {\bibfnamefont {F.}~\bibnamefont
  {Caglieris}}, \bibinfo {author} {\bibfnamefont {A.}~\bibnamefont {Braggio}},
  \bibinfo {author} {\bibfnamefont {I.}~\bibnamefont {Pallecchi}}, \bibinfo
  {author} {\bibfnamefont {A.}~\bibnamefont {Provino}}, \bibinfo {author}
  {\bibfnamefont {M.}~\bibnamefont {Pani}}, \bibinfo {author} {\bibfnamefont
  {G.}~\bibnamefont {Lamura}}, \bibinfo {author} {\bibfnamefont
  {A.}~\bibnamefont {Jost}}, \bibinfo {author} {\bibfnamefont {U.}~\bibnamefont
  {Zeitler}}, \bibinfo {author} {\bibfnamefont {E.}~\bibnamefont
  {Galleani~D'Agliano}}, \bibinfo {author} {\bibfnamefont {P.}~\bibnamefont
  {Manfrinetti}}, \ and\ \bibinfo {author} {\bibfnamefont {M.}~\bibnamefont
  {Putti}},\ }\bibfield  {title} {\enquote {\bibinfo {title} {Magneto-Seebeck
  effect in $r\mathrm{FeAsO}$ ($r=\mathrm{rare}$ earth) compounds: Probing the
  magnon drag scenario},}\ }\href {\doibase 10.1103/PhysRevB.90.134421} {\bibfield  {journal}
  {\bibinfo  {journal} {Phys. Rev. B}\ }\textbf {\bibinfo {volume} {90}},\
  \bibinfo {pages} {134421} (\bibinfo {year} {2014})}\BibitemShut {NoStop}%
\bibitem [{\citenamefont {Germanese}\ \emph
  {et~al.}({\natexlab{a}})\citenamefont {Germanese}, \citenamefont {Paolucci},
  \citenamefont {Marchegiani}, \citenamefont {Braggio},\ and\ \citenamefont
  {Giazotto}}]{Patent1}%
  \BibitemOpen
  \bibfield  {author} {\bibinfo {author} {\bibfnamefont {G.}~\bibnamefont
  {Germanese}}, \bibinfo {author} {\bibfnamefont {F.}~\bibnamefont {Paolucci}},
  \bibinfo {author} {\bibfnamefont {G.}~\bibnamefont {Marchegiani}}, \bibinfo
  {author} {\bibfnamefont {A.}~\bibnamefont {Braggio}}, \ and\ \bibinfo
  {author} {\bibfnamefont {F.}~\bibnamefont {Giazotto}},\ }\href@noop {}
  {\enquote {\bibinfo {title} {Superconducting bipolar thermoelectric memory
  and method for writing a superconducting bipolar thermo-electric memory},}\ }
   \bibinfo {note} {I.T. Patent, 102021000032042
  (2021)}\BibitemShut {NoStop}%
\bibitem [{\citenamefont {Germanese}\ \emph
  {et~al.}({\natexlab{b}})\citenamefont {Germanese}, \citenamefont {Paolucci},
  \citenamefont {Braggio},\ and\ \citenamefont {Giazotto}}]{Patent2}%
  \BibitemOpen
  \bibfield  {author} {\bibinfo {author} {\bibfnamefont {G.}~\bibnamefont
  {Germanese}}, \bibinfo {author} {\bibfnamefont {F.}~\bibnamefont {Paolucci}},
  \bibinfo {author} {\bibfnamefont {A.}~\bibnamefont {Braggio}}, \ and\
  \bibinfo {author} {\bibfnamefont {F.}~\bibnamefont {Giazotto}},\ }\href@noop
  {} {\enquote {\bibinfo {title} {Broadband passive superconducting
  thermoelectric single photon-detector},}\ }  \bibinfo
  {note} {I.T. Patent, 102023000001854 (2023)}\BibitemShut {NoStop}%
\bibitem [{\citenamefont {Benenti}\ \emph {et~al.}(2017)\citenamefont
  {Benenti}, \citenamefont {Casati}, \citenamefont {Saito},\ and\ \citenamefont
  {Whitney}}]{Ben17}%
  \BibitemOpen
  \bibfield  {author} {\bibinfo {author} {\bibfnamefont {G.}~\bibnamefont
  {Benenti}}, \bibinfo {author} {\bibfnamefont {G.}~\bibnamefont {Casati}},
  \bibinfo {author} {\bibfnamefont {K.}~\bibnamefont {Saito}}, \ and\ \bibinfo
  {author} {\bibfnamefont {R.}~\bibnamefont {Whitney}},\ }\bibfield  {title} {\enquote {\bibinfo {title}
  {Fundamental aspects of steady-state conversion of heat to work at the   nanoscale},}\ }\href {\doibase
  https://doi.org/10.1016/j.physrep.2017.05.008} {\bibfield  {journal}
  {\bibinfo  {journal} {Phys. Rep.}\ }\textbf {\bibinfo {volume} {694}},\
  \bibinfo {pages} {1} (\bibinfo {year} {2017})}\BibitemShut {NoStop}%
\bibitem [{\citenamefont {Dynes}\ \emph {et~al.}(1978)\citenamefont {Dynes},
  \citenamefont {Narayanamurti},\ and\ \citenamefont {Garno}}]{Dyn78}%
  \BibitemOpen
  \bibfield  {author} {\bibinfo {author} {\bibfnamefont {R.~C.}\ \bibnamefont
  {Dynes}}, \bibinfo {author} {\bibfnamefont {V.}~\bibnamefont
  {Narayanamurti}}, \ and\ \bibinfo {author} {\bibfnamefont {J.~P.}\
  \bibnamefont {Garno}},\ }\bibfield  {title} {\enquote {\bibinfo {title}
  {Direct measurement of quasiparticle-lifetime broadening in a strong-coupled
  superconductor},}\ }\href {\doibase 10.1103/PhysRevLett.41.1509}
  {\bibfield  {journal} {\bibinfo  {journal} {Phys. Rev. Lett.}\ }\textbf
  {\bibinfo {volume} {41}},\ \bibinfo {pages} {1509} (\bibinfo {year}
  {1978})}\BibitemShut {NoStop}%
\bibitem [{\citenamefont {Guarcello}\ \emph
  {et~al.}(2019{\natexlab{a}})\citenamefont {Guarcello}, \citenamefont
  {Braggio}, \citenamefont {Solinas},\ and\ \citenamefont
  {Giazotto}}]{GuaBra19}%
  \BibitemOpen
  \bibfield  {author} {\bibinfo {author} {\bibfnamefont {C.}~\bibnamefont
  {Guarcello}}, \bibinfo {author} {\bibfnamefont {A.}~\bibnamefont {Braggio}},
  \bibinfo {author} {\bibfnamefont {P.}~\bibnamefont {Solinas}}, \ and\
  \bibinfo {author} {\bibfnamefont {F.}~\bibnamefont {Giazotto}},\ }\bibfield  {title}
  {\enquote {\bibinfo {title} {Nonlinear critical-current thermal response of
  an asymmetric Josephson tunnel junction},}\ }\href
  {\doibase 10.1103/PhysRevApplied.11.024002} {\bibfield  {journal} {\bibinfo
  {journal} {Phys. Rev. Applied}\ }\textbf {\bibinfo {volume} {11}},\ \bibinfo
  {pages} {024002} (\bibinfo {year} {2019}{\natexlab{a}})}\BibitemShut
  {NoStop}%
\bibitem [{\citenamefont {Guarcello}\ \emph
  {et~al.}(2019{\natexlab{b}})\citenamefont {Guarcello}, \citenamefont
  {Braggio}, \citenamefont {Solinas}, \citenamefont {Pepe},\ and\ \citenamefont
  {Giazotto}}]{Gua19}%
  \BibitemOpen
  \bibfield  {author} {\bibinfo {author} {\bibfnamefont {C.}~\bibnamefont
  {Guarcello}}, \bibinfo {author} {\bibfnamefont {A.}~\bibnamefont {Braggio}},
  \bibinfo {author} {\bibfnamefont {P.}~\bibnamefont {Solinas}}, \bibinfo
  {author} {\bibfnamefont {G.~P.}\ \bibnamefont {Pepe}}, \ and\ \bibinfo
  {author} {\bibfnamefont {F.}~\bibnamefont {Giazotto}},\ }\bibfield  {title}
  {\enquote {\bibinfo {title} {Josephson-threshold calorimeter},}\ }\href {\doibase
  10.1103/PhysRevApplied.11.054074} {\bibfield  {journal} {\bibinfo  {journal}
  {Phys. Rev. Applied}\ }\textbf {\bibinfo {volume} {11}},\ \bibinfo {pages}
  {054074} (\bibinfo {year} {2019}{\natexlab{b}})}\BibitemShut {NoStop}%
\bibitem [{Note1()}]{Note1}%
  \BibitemOpen
  \bibinfo {note} {This can be done using different strategies such as
  increasing the barrier opacity or using SQUID-like interference, or even
  Fraunhofer-like suppression~\cite
  {Marchegiani2020,Germanese2022}.}\BibitemShut {Stop}%
\bibitem [{\citenamefont {Raghu}\ \emph {et~al.}(2008)\citenamefont {Raghu},
  \citenamefont {Qi}, \citenamefont {Liu}, \citenamefont {Scalapino},\ and\
  \citenamefont {Zhang}}]{Rag08}%
  \BibitemOpen
  \bibfield  {author} {\bibinfo {author} {\bibfnamefont {S.}~\bibnamefont
  {Raghu}}, \bibinfo {author} {\bibfnamefont {X.-L.}\ \bibnamefont {Qi}},
  \bibinfo {author} {\bibfnamefont {C.-X.}\ \bibnamefont {Liu}}, \bibinfo
  {author} {\bibfnamefont {D.~J.}\ \bibnamefont {Scalapino}}, \ and\ \bibinfo
  {author} {\bibfnamefont {S.-C.}\ \bibnamefont {Zhang}},\ }\bibfield
  {title} {\enquote {\bibinfo {title} {Minimal two-band model of the
  superconducting iron oxypnictides},}\ }\href {\doibase
  10.1103/PhysRevB.77.220503} {\bibfield  {journal} {\bibinfo  {journal} {Phys.
  Rev. B}\ }\textbf {\bibinfo {volume} {77}},\ \bibinfo {pages} {220503}
  (\bibinfo {year} {2008})}\BibitemShut {NoStop}%
\bibitem [{Note2()}]{Note2}%
  \BibitemOpen
  \bibinfo {note} {Our goal is to understand the influence of multi-band
  superconductivity on TE properties, thus Raghu's approach is sufficient,
  although we are aware that models with more orbitals have been also
  developed~\cite {Esc09,Nic17}.often used to investigate multi-band effects on
  superconductivity and magnetism in FeSC materials~\cite
  {Que16,Cav21}~}\BibitemShut {NoStop}%
\bibitem [{\citenamefont {Parish}\ \emph {et~al.}(2008)\citenamefont {Parish},
  \citenamefont {Hu},\ and\ \citenamefont {Bernevig}}]{Par08}%
  \BibitemOpen
  \bibfield  {author} {\bibinfo {author} {\bibfnamefont {M.~M.}\ \bibnamefont
  {Parish}}, \bibinfo {author} {\bibfnamefont {J.}~\bibnamefont {Hu}}, \ and\
  \bibinfo {author} {\bibfnamefont {B.~A.}\ \bibnamefont {Bernevig}},\ }\bibfield  {title} {\enquote {\bibinfo {title}
  {Experimental consequences of the $s$-wave
  $\text{cos}({k}_{x})\text{cos}({k}_{y})$ superconductivity in the iron
  pnictides},}\ }\href
  {\doibase 10.1103/PhysRevB.78.144514} {\bibfield  {journal} {\bibinfo
  {journal} {Phys. Rev. B}\ }\textbf {\bibinfo {volume} {78}},\ \bibinfo
  {pages} {144514} (\bibinfo {year} {2008})}\BibitemShut {NoStop}%
    \bibitem [{NoteSM()}]{NoteSM}%
  \BibitemOpen
  \bibinfo {note} {See Supplemental Material at \href{https://journals.aps.org/prb/supplemental/10.1103/PhysRevB.108.L100511}{https://journals.aps.org/prb/supplemental/10.1103/Phys RevB.108.L100511} for the details on the model used for describing FeSCs; details on the FeSC DoS; the calculation of linear TE coefficients in the cases discussed in Fig.~\ref{Figure02}; the calculation of TE effects in the case of other OPSs; the robustness of TE effects to chemical potential variations; the calculation of the power factor; the relation between the superconducting gaps at the optimal TE efficiency. The Supplemental Material also contains Refs.~\cite{Yam13,Dum16,Liu18}.}\BibitemShut {Stop}%
\bibitem [{\citenamefont {Bang}\ and\ \citenamefont {Stewart}(2017)}]{Ban17}%
  \BibitemOpen
  \bibfield  {author} {\bibinfo {author} {\bibfnamefont {Y.}~\bibnamefont
  {Bang}}\ and\ \bibinfo {author} {\bibfnamefont {G.~R.}\ \bibnamefont
  {Stewart}},\ }\bibfield  {title} {\enquote {\bibinfo {title} {Superconducting
  properties of the $s^\pm$-wave state: Fe-based superconductors},}\ }\href {\doibase 10.1088/1361-648X/aa564b} {\bibfield  {journal}
  {\bibinfo  {journal} {J. Phys.: Condens. Matter}\ }\textbf {\bibinfo {volume}
  {29}},\ \bibinfo {pages} {123003} (\bibinfo {year} {2017})}\BibitemShut
  {NoStop}%
\bibitem [{\citenamefont {Fernandes}\ \emph {et~al.}(2022)\citenamefont
  {Fernandes}, \citenamefont {Coldea}, \citenamefont {Ding}, \citenamefont
  {Fisher}, \citenamefont {Hirschfeld},\ and\ \citenamefont {Kotliar}}]{Fer22}%
  \BibitemOpen
  \bibfield  {author} {\bibinfo {author} {\bibfnamefont {R.~M.}\ \bibnamefont
  {Fernandes}}, \bibinfo {author} {\bibfnamefont {A.~I.}\ \bibnamefont
  {Coldea}}, \bibinfo {author} {\bibfnamefont {H.}~\bibnamefont {Ding}},
  \bibinfo {author} {\bibfnamefont {I.~R.}\ \bibnamefont {Fisher}}, \bibinfo
  {author} {\bibfnamefont {P.~J.}\ \bibnamefont {Hirschfeld}}, \ and\ \bibinfo
  {author} {\bibfnamefont {G.}~\bibnamefont {Kotliar}},\ }\bibfield  {title} {\enquote {\bibinfo {title} {Iron pnictides and
  chalcogenides: a new paradigm for superconductivity},}\ }\href {\doibase
  10.1038/s41586-021-04073-2} {\bibfield  {journal} {\bibinfo  {journal}
  {Nature}\ }\textbf {\bibinfo {volume} {601}},\ \bibinfo {pages} {35}
  (\bibinfo {year} {2022})}\BibitemShut {NoStop}%
\bibitem [{Note3()}]{Note3}%
  \BibitemOpen
  \bibinfo {note} {Hereafter, we assume an interorbital hopping parameter
  $\left | t_1 \right |=0.15\protect \tmspace +\thickmuskip {.2777em}\protect
  \text {eV}$, in line with the value often used in literature, e.g.,
  Ref.~\cite {Wan15}.}\BibitemShut {Stop}%
\bibitem [{\citenamefont {Ptok}(2014)}]{Pto14}%
  \BibitemOpen
  \bibfield  {author} {\bibinfo {author} {\bibfnamefont {A.}~\bibnamefont
  {Ptok}},\ }\bibfield  {title} {\enquote {\bibinfo {title} {Influence of
  s$_\pm$ symmetry on unconventional superconductivity in pnictides above
  the Pauli limit -- two-band model study},}\ }\href {\doibase 10.1140/epjb/e2013-41007-2} {\bibfield  {journal}
  {\bibinfo  {journal} {Eur. Phys. J. B}\ }\textbf {\bibinfo {volume} {87}},\
  \bibinfo {pages} {2} (\bibinfo {year} {2014})}\BibitemShut {NoStop}%
\bibitem [{Note3.5()}]{Note3.5}%
  \BibitemOpen
  \bibinfo {note} {For a complete TE characterization of the device, one should require also to look at at the power that the system produces as an energy harvester, which is quantified by the \emph{power factor}, $\text{PF}=\sigma\,S^2$~\cite{Ben17}. More details and the specific calculation of $\text{PF}$ in the cases of interest for the present work are reported in Ref~\cite{NoteSM}.}\BibitemShut
  {Stop}%
  \bibitem [{Note4()}]{Note4}%
  \BibitemOpen
  \bibinfo {note} {FeSCs have been found to exhibit a wide range of
  superconducting transition temperatures~\cite {Fer22}, albeit that the
  $2\Delta _{max}/(k_BT_c)$ ratio (with $\Delta _{max}$ being the
  zero-temperature value of the largest gap) often falls within $6.0-8.5$, in
  contrast to the $\sim 3.5$ value of conventional BCS SCs.}\BibitemShut
  {Stop}%
\bibitem [{Note5()}]{Note5}%
  \BibitemOpen
  \bibinfo {note} {In this work we are assuming a temperature-independent gap
  because we typically consider temperatures $T<0.4T_c$ since the FeSC
  superconductive gap shows a BCS-like temperature dependence~\cite
  {Jin10}.}\BibitemShut {Stop}%
\bibitem [{Note6()}]{Note6}%
  \BibitemOpen
  \bibinfo {note} {The value $\Delta _0^{\protect \text {FeSC}}=0.1$, hereafter
  used in the manuscript, corresponds to $T_c\in [41-58]\protect \tmspace
  +\thickmuskip {.2777em}\protect \text {K}$.}\BibitemShut {Stop}%
\bibitem [{\citenamefont {Walter}\ \emph {et~al.}(2011)\citenamefont {Walter},
  \citenamefont {Walowski}, \citenamefont {Zbarsky}, \citenamefont
  {M{\"u}nzenberg}, \citenamefont {Sch{\"a}fers}, \citenamefont {Ebke},
  \citenamefont {Reiss}, \citenamefont {Thomas}, \citenamefont {Peretzki},
  \citenamefont {Seibt}, \citenamefont {Moodera}, \citenamefont {Czerner},
  \citenamefont {Bachmann},\ and\ \citenamefont {Heiliger}}]{Wal11}%
  \BibitemOpen
  \bibfield  {author} {\bibinfo {author} {\bibfnamefont {M.}~\bibnamefont
  {Walter}}, \bibinfo {author} {\bibfnamefont {J.}~\bibnamefont {Walowski}},
  \bibinfo {author} {\bibfnamefont {V.}~\bibnamefont {Zbarsky}}, \bibinfo
  {author} {\bibfnamefont {M.}~\bibnamefont {M{\"u}nzenberg}}, \bibinfo
  {author} {\bibfnamefont {M.}~\bibnamefont {Sch{\"a}fers}}, \bibinfo {author}
  {\bibfnamefont {D.}~\bibnamefont {Ebke}}, \bibinfo {author} {\bibfnamefont
  {G.}~\bibnamefont {Reiss}}, \bibinfo {author} {\bibfnamefont
  {A.}~\bibnamefont {Thomas}}, \bibinfo {author} {\bibfnamefont
  {P.}~\bibnamefont {Peretzki}}, \bibinfo {author} {\bibfnamefont
  {M.}~\bibnamefont {Seibt}}, \bibinfo {author} {\bibfnamefont {J.~S.}\
  \bibnamefont {Moodera}}, \bibinfo {author} {\bibfnamefont {M.}~\bibnamefont
  {Czerner}}, \bibinfo {author} {\bibfnamefont {M.}~\bibnamefont {Bachmann}}, \
  and\ \bibinfo {author} {\bibfnamefont {C.}~\bibnamefont {Heiliger}},\ }\bibfield  {title}
  {\enquote {\bibinfo {title} {Seebeck effect in magnetic tunnel junctions},}\
  }\href
  {\doibase 10.1038/nmat3076} {\bibfield  {journal} {\bibinfo  {journal} {Nat.
  Mater.}\ }\textbf {\bibinfo {volume} {10}},\ \bibinfo {pages} {742} (\bibinfo
  {year} {2011})}\BibitemShut {NoStop}%
\bibitem [{\citenamefont {Kolenda}\ \emph
  {et~al.}(2016{\natexlab{b}})\citenamefont {Kolenda}, \citenamefont {Wolf},\
  and\ \citenamefont {Beckmann}}]{Kol16}%
  \BibitemOpen
  \bibfield  {author} {\bibinfo {author} {\bibfnamefont {S.}~\bibnamefont
  {Kolenda}}, \bibinfo {author} {\bibfnamefont {M.~J.}\ \bibnamefont {Wolf}}, \
  and\ \bibinfo {author} {\bibfnamefont {D.}~\bibnamefont {Beckmann}},\ }\bibfield  {title} {\enquote {\bibinfo {title} {Observation of
  thermoelectric currents in high-field superconductor-ferromagnet tunnel
  junctions},}\ }\href
  {\doibase 10.1103/PhysRevLett.116.097001} {\bibfield  {journal} {\bibinfo
  {journal} {Phys. Rev. Lett.}\ }\textbf {\bibinfo {volume} {116}},\ \bibinfo
  {pages} {097001} (\bibinfo {year} {2016}{\natexlab{b}})}\BibitemShut
  {NoStop}%
\bibitem [{\citenamefont {Gonz\'alez-Ruano}\ \emph {et~al.}(2023)\citenamefont
  {Gonz\'alez-Ruano}, \citenamefont {Caso}, \citenamefont {Ouassou},
  \citenamefont {Tiusan}, \citenamefont {Lu}, \citenamefont {Linder},\ and\
  \citenamefont {Aliev}}]{Gon23}%
  \BibitemOpen
  \bibfield  {author} {\bibinfo {author} {\bibfnamefont {C.}~\bibnamefont
  {Gonz\'alez-Ruano}}, \bibinfo {author} {\bibfnamefont {D.}~\bibnamefont
  {Caso}}, \bibinfo {author} {\bibfnamefont {J.~A.}\ \bibnamefont {Ouassou}},
  \bibinfo {author} {\bibfnamefont {C.}~\bibnamefont {Tiusan}}, \bibinfo
  {author} {\bibfnamefont {Y.}~\bibnamefont {Lu}}, \bibinfo {author}
  {\bibfnamefont {J.}~\bibnamefont {Linder}}, \ and\ \bibinfo {author}
  {\bibfnamefont {F.~G.}\ \bibnamefont {Aliev}},\ }\bibfield  {title} {\enquote {\bibinfo {title} {Observation of Magnetic State Dependent Thermoelectricity in Superconducting Spin Valves},}\ }\href {\doibase
  10.1103/PhysRevLett.130.237001} {\bibfield  {journal} {\bibinfo  {journal}
  {Phys. Rev. Lett.}\ }\textbf {\bibinfo {volume} {130}},\ \bibinfo {pages}
  {237001} (\bibinfo {year} {2023})}\BibitemShut {NoStop}%
\bibitem [{\citenamefont {Svilans}\ \emph {et~al.}(2016)\citenamefont
  {Svilans}, \citenamefont {Leijnse},\ and\ \citenamefont {Linke}}]{Svi16}%
  \BibitemOpen
  \bibfield  {author} {\bibinfo {author} {\bibfnamefont {A.}~\bibnamefont
  {Svilans}}, \bibinfo {author} {\bibfnamefont {M.}~\bibnamefont {Leijnse}}, \
  and\ \bibinfo {author} {\bibfnamefont {H.}~\bibnamefont {Linke}},\ }\bibfield  {title} {\enquote {\bibinfo {title} {Experiments on the thermoelectric properties of quantum dots},}\ }\href
  {\doibase https://doi.org/10.1016/j.crhy.2016.08.002} {\bibfield  {journal}
  {\bibinfo  {journal} {C. R. Phys.}\ }\textbf {\bibinfo {volume}
  {17}},\ \bibinfo {pages} {1096} (\bibinfo {year} {2016})}\BibitemShut
  {NoStop}%
\bibitem [{Note7()}]{Note7}%
  \BibitemOpen
  \bibinfo {note} {We included in Fig.~\ref {Figure02}(e-f) a black dashed line to mark the condition $T=T_c$ at which the the non-iron electrode ceases to be superconductive and the system actually behaves as an FeSC-I-N junction. Figure~\ref {Figure02}(f) allows also to investigate the functional dependence existing between the superconducting gaps in the region of parameter space that give the optimal TE efficiency~\cite{NoteSM}.}\BibitemShut {Stop}%
\bibitem [{Note8()}]{Note8}%
  \BibitemOpen
  \bibinfo {note} {The Raghu's approach entails two pairing gaps, one for each
  orbital, $\Delta _{1,2}$, that satisfy the condition $\Delta
  _1(k_x,k_y)=\Delta _2(k_y,k_x)$ for all the pairing symmetries described
  above, except for $d_{x^2-y^2}$ giving $\Delta _1(k_x,k_y)=-\Delta
  _2(k_x,k_y)$~\cite {Par08,Seo08}. In the latter case, the eigenvalues and the
  DoS are not reduced to the simple expressions given in Eqs.~\protect \textup
  {\hbox {\mathsurround \z@ \protect \normalfont (\ignorespaces \ref
  {eigenvalues}\unskip \@@italiccorr )}}$-$\protect \textup {\hbox
  {\mathsurround \z@ \protect \normalfont (\ignorespaces \ref
  {eq:IBDoSpm}\unskip \@@italiccorr )}}, but we use the general expression of
  the spectral function given in Ref.~\cite {Par08} with the eigenvalues
  presented in Ref~\cite {Seo08}.}\BibitemShut {Stop}%
\bibitem [{\citenamefont {Seo}\ \emph {et~al.}(2008)\citenamefont {Seo},
  \citenamefont {Bernevig},\ and\ \citenamefont {Hu}}]{Seo08}%
  \BibitemOpen
  \bibfield  {author} {\bibinfo {author} {\bibfnamefont {K.}~\bibnamefont
  {Seo}}, \bibinfo {author} {\bibfnamefont {B.~A.}\ \bibnamefont {Bernevig}}, \
  and\ \bibinfo {author} {\bibfnamefont {J.}~\bibnamefont {Hu}},\ }\bibfield  {title} {\enquote {\bibinfo {title} {Pairing
  symmetry in a two-orbital exchange coupling model of oxypnictides},}\ }\href
  {\doibase 10.1103/PhysRevLett.101.206404} {\bibfield  {journal} {\bibinfo
  {journal} {Phys. Rev. Lett.}\ }\textbf {\bibinfo {volume} {101}},\ \bibinfo
  {pages} {206404} (\bibinfo {year} {2008})}\BibitemShut {NoStop}%
  \bibitem [{\citenamefont {Eschrig}\ and\ \citenamefont
  {Koepernik}(2009)}]{Esc09}%
  \BibitemOpen
  \bibfield  {author} {\bibinfo {author} {\bibfnamefont {H.}~\bibnamefont
  {Eschrig}}\ and\ \bibinfo {author} {\bibfnamefont {K.}~\bibnamefont
  {Koepernik}},\ }\bibfield  {title} {\enquote {\bibinfo {title} {Tight-binding
  models for the iron-based superconductors},}\ }\href {\doibase 10.1103/PhysRevB.80.104503} {\bibfield
  {journal} {\bibinfo  {journal} {Phys. Rev. B}\ }\textbf {\bibinfo {volume}
  {80}},\ \bibinfo {pages} {104503} (\bibinfo {year} {2009})}\BibitemShut
  {NoStop}%
\bibitem [{\citenamefont {Nica}\ \emph {et~al.}(2017)\citenamefont {Nica},
  \citenamefont {Yu},\ and\ \citenamefont {Si}}]{Nic17}%
  \BibitemOpen
  \bibfield  {author} {\bibinfo {author} {\bibfnamefont {E.~M.}\ \bibnamefont
  {Nica}}, \bibinfo {author} {\bibfnamefont {R.}~\bibnamefont {Yu}}, \ and\
  \bibinfo {author} {\bibfnamefont {Q.}~\bibnamefont {Si}},\ }\bibfield  {title} {\enquote {\bibinfo {title} {Orbital-selective pairing
  and superconductivity in iron selenides},}\ }\href {\doibase
  10.1038/s41535-017-0027-6} {\bibfield  {journal} {\bibinfo  {journal} {npj
  Quantum Mater.}\ }\textbf {\bibinfo {volume} {2}},\ \bibinfo {pages} {24}
  (\bibinfo {year} {2017})}\BibitemShut {NoStop}%
\bibitem [{\citenamefont {{Querales Flores}}\ \emph {et~al.}(2016)\citenamefont
  {{Querales Flores}}, \citenamefont {Ventura}, \citenamefont {Citro},\ and\
  \citenamefont {Rodríguez-Núñez}}]{Que16}%
  \BibitemOpen
  \bibfield  {author} {\bibinfo {author} {\bibfnamefont {J.}~\bibnamefont
  {{Querales Flores}}}, \bibinfo {author} {\bibfnamefont {C.}~\bibnamefont
  {Ventura}}, \bibinfo {author} {\bibfnamefont {R.}~\bibnamefont {Citro}}, \
  and\ \bibinfo {author} {\bibfnamefont {J.}~\bibnamefont
  {Rodr\'iguez-N\'u$\tilde{\text{n}}$ez}},\ }\bibfield  {title} {\enquote {\bibinfo {title}
  {Temperature and doping dependence of normal state spectral properties in a
  two-orbital model for ferropnictides},}\ }\href {\doibase
  https://doi.org/10.1016/j.physleta.2016.02.042} {\bibfield  {journal}
  {\bibinfo  {journal} {Phys. Lett. A}\ }\textbf {\bibinfo {volume} {380}},\
  \bibinfo {pages} {1648} (\bibinfo {year} {2016})}\BibitemShut {NoStop}%
\bibitem [{\citenamefont {Cavanagh}\ and\ \citenamefont
  {Brydon}(2021)}]{Cav21}%
  \BibitemOpen
  \bibfield  {author} {\bibinfo {author} {\bibfnamefont {D.~C.}\ \bibnamefont
  {Cavanagh}}\ and\ \bibinfo {author} {\bibfnamefont {P.~M.~R.}\ \bibnamefont
  {Brydon}},\ }\bibfield  {title} {\enquote {\bibinfo {title} {General theory
  of robustness against disorder in multiband superconductors},}\ }\href {\doibase 10.1103/PhysRevB.104.014503} {\bibfield
  {journal} {\bibinfo  {journal} {Phys. Rev. B}\ }\textbf {\bibinfo {volume}
  {104}},\ \bibinfo {pages} {014503} (\bibinfo {year} {2021})}\BibitemShut
  {NoStop}%
\bibitem [{\citenamefont {Wang}\ and\ \citenamefont
  {Nevidomskyy}(2015)}]{Wan15}%
  \BibitemOpen
  \bibfield  {author} {\bibinfo {author} {\bibfnamefont {Z.}~\bibnamefont
  {Wang}}\ and\ \bibinfo {author} {\bibfnamefont {A.~H.}\ \bibnamefont
  {Nevidomskyy}},\ }\bibfield  {title} {\enquote {\bibinfo {title} {Orbital
  nematic order and interplay with magnetism in the two-orbital hubbard
  model},}\ }\href {\doibase 10.1088/0953-8984/27/22/225602} {\bibfield
  {journal} {\bibinfo  {journal} {J. Phys.: Condens. Matter}\ }\textbf
  {\bibinfo {volume} {27}},\ \bibinfo {pages} {225602} (\bibinfo {year}
  {2015})}\BibitemShut {NoStop}%
\bibitem [{\citenamefont {Jin}\ \emph {et~al.}(2010)\citenamefont {Jin},
  \citenamefont {Pan}, \citenamefont {He}, \citenamefont {Li}, \citenamefont
  {Li}, \citenamefont {wen Peng}, \citenamefont {Thompson}, \citenamefont
  {Sales}, \citenamefont {Sefat}, \citenamefont {McGuire}, \citenamefont
  {Mandrus}, \citenamefont {Wendelken}, \citenamefont {Keppens},\ and\
  \citenamefont {Plummer}}]{Jin10}%
  \BibitemOpen
  \bibfield  {author} {\bibinfo {author} {\bibfnamefont {R.}~\bibnamefont
  {Jin}}, \bibinfo {author} {\bibfnamefont {M.~H.}\ \bibnamefont {Pan}},
  \bibinfo {author} {\bibfnamefont {X.~B.}\ \bibnamefont {He}}, \bibinfo
  {author} {\bibfnamefont {G.}~\bibnamefont {Li}}, \bibinfo {author}
  {\bibfnamefont {D.}~\bibnamefont {Li}}, \bibinfo {author} {\bibfnamefont
  {R.}~\bibnamefont {wen Peng}}, \bibinfo {author} {\bibfnamefont {J.~R.}\
  \bibnamefont {Thompson}}, \bibinfo {author} {\bibfnamefont {B.~C.}\
  \bibnamefont {Sales}}, \bibinfo {author} {\bibfnamefont {A.~S.}\ \bibnamefont
  {Sefat}}, \bibinfo {author} {\bibfnamefont {M.~A.}\ \bibnamefont {McGuire}},
  \bibinfo {author} {\bibfnamefont {D.}~\bibnamefont {Mandrus}}, \bibinfo
  {author} {\bibfnamefont {J.~F.}\ \bibnamefont {Wendelken}}, \bibinfo {author}
  {\bibfnamefont {V.}~\bibnamefont {Keppens}}, \ and\ \bibinfo {author}
  {\bibfnamefont {E.~W.}\ \bibnamefont {Plummer}},\ }\bibfield  {title} {\enquote
  {\bibinfo {title} {Electronic, magnetic and optical properties of two
  Fe-based superconductors and related parent compounds},}\ }\href {\doibase
  10.1088/0953-2048/23/5/054005} {\bibfield  {journal} {\bibinfo  {journal}
  {Supercond. Sci. Technol.}\ }\textbf {\bibinfo {volume} {23}},\ \bibinfo
  {pages} {054005} (\bibinfo {year} {2010})}\BibitemShut {NoStop}%
\bibitem [{\citenamefont {Yamase}\ and\ \citenamefont {Zeyher}(2013)}]{Yam13}%
  \BibitemOpen
  \bibfield  {author} {\bibinfo {author} {\bibfnamefont {H.}~\bibnamefont
  {Yamase}}\ and\ \bibinfo {author} {\bibfnamefont {R.}~\bibnamefont
  {Zeyher}},\ }\bibfield  {title} {\enquote
  {\bibinfo {title} {Superconductivity from orbital nematic fluctuations},}\ }\href {\doibase 10.1103/PhysRevB.88.180502} {\bibfield
  {journal} {\bibinfo  {journal} {Phys. Rev. B}\ }\textbf {\bibinfo {volume}
  {88}},\ \bibinfo {pages} {180502} (\bibinfo {year} {2013})}\BibitemShut
  {NoStop}%
\bibitem [{\citenamefont {Dumitrescu}\ \emph {et~al.}(2016)\citenamefont
  {Dumitrescu}, \citenamefont {Serbyn}, \citenamefont {Scalettar},\ and\
  \citenamefont {Vishwanath}}]{Dum16}%
  \BibitemOpen
  \bibfield  {author} {\bibinfo {author} {\bibfnamefont {P.~T.}\ \bibnamefont
  {Dumitrescu}}, \bibinfo {author} {\bibfnamefont {M.}~\bibnamefont {Serbyn}},
  \bibinfo {author} {\bibfnamefont {R.~T.}\ \bibnamefont {Scalettar}}, \ and\
  \bibinfo {author} {\bibfnamefont {A.}~\bibnamefont {Vishwanath}},\ }\bibfield  {title} {\enquote
  {\bibinfo {title} {Superconductivity and nematic fluctuations in a model of doped FeSe monolayers: Determinant quantum Monte Carlo study},}\ }\href
  {\doibase 10.1103/PhysRevB.94.155127} {\bibfield  {journal} {\bibinfo
  {journal} {Phys. Rev. B}\ }\textbf {\bibinfo {volume} {94}},\ \bibinfo
  {pages} {155127} (\bibinfo {year} {2016})}\BibitemShut {NoStop}%
  \bibitem [{\citenamefont {Liu}\ \emph {et~al.}(2018)\citenamefont {Liu},
  \citenamefont {Fang}, \citenamefont {Zheng}, \citenamefont {Huang},\ and\
  \citenamefont {Lin}}]{Liu18}%
  \BibitemOpen
  \bibfield  {author} {\bibinfo {author} {\bibfnamefont {G.}~\bibnamefont
  {Liu}}, \bibinfo {author} {\bibfnamefont {S.}~\bibnamefont {Fang}}, \bibinfo
  {author} {\bibfnamefont {X.}~\bibnamefont {Zheng}}, \bibinfo {author}
  {\bibfnamefont {Z.}~\bibnamefont {Huang}}, \ and\ \bibinfo {author}
  {\bibfnamefont {H.}~\bibnamefont {Lin}},\ }\bibfield  {title} {\enquote
  {\bibinfo {title} {Interplay between nematic fluctuation and superconductivity in a two-orbital Hubbard model: a quantum Monte Carlo study},}\ }\href {\doibase
  10.1088/1361-648x/aae289} {\bibfield  {journal} {\bibinfo  {journal} {J.
  Phys.: Condens. Matter}\ }\textbf {\bibinfo {volume} {30}},\ \bibinfo {pages}
  {445604} (\bibinfo {year} {2018})}\BibitemShut {NoStop}%
\end{thebibliography}

\begin{thebibliography}{9}%
\makeatletter
\providecommand \@ifxundefined [1]{%
 \@ifx{#1\undefined}
}%
\providecommand \@ifnum [1]{%
 \ifnum #1\expandafter \@firstoftwo
 \else \expandafter \@secondoftwo
 \fi
}%
\providecommand \@ifx [1]{%
 \ifx #1\expandafter \@firstoftwo
 \else \expandafter \@secondoftwo
 \fi
}%
\providecommand \natexlab [1]{#1}%
\providecommand \enquote  [1]{``#1''}%
\providecommand \bibnamefont  [1]{#1}%
\providecommand \bibfnamefont [1]{#1}%
\providecommand \citenamefont [1]{#1}%
\providecommand \href@noop [0]{\@secondoftwo}%
\providecommand \href [0]{\begingroup \@sanitize@url \@href}%
\providecommand \@href[1]{\@@startlink{#1}\@@href}%
\providecommand \@@href[1]{\endgroup#1\@@endlink}%
\providecommand \@sanitize@url [0]{\catcode `\\12\catcode `\$12\catcode
  `\&12\catcode `\#12\catcode `\^12\catcode `\_12\catcode `\%12\relax}%
\providecommand \@@startlink[1]{}%
\providecommand \@@endlink[0]{}%
\providecommand \url  [0]{\begingroup\@sanitize@url \@url }%
\providecommand \@url [1]{\endgroup\@href {#1}{\urlprefix }}%
\providecommand \urlprefix  [0]{URL }%
\providecommand \Eprint [0]{\href }%
\providecommand \doibase [0]{http://dx.doi.org/}%
\providecommand \selectlanguage [0]{\@gobble}%
\providecommand \bibinfo  [0]{\@secondoftwo}%
\providecommand \bibfield  [0]{\@secondoftwo}%
\providecommand \translation [1]{[#1]}%
\providecommand \BibitemOpen [0]{}%
\providecommand \bibitemStop [0]{}%
\providecommand \bibitemNoStop [0]{.\EOS\space}%
\providecommand \EOS [0]{\spacefactor3000\relax}%
\providecommand \BibitemShut  [1]{\csname bibitem#1\endcsname}%
\let\auto@bib@innerbib\@empty
\bibitem [{\citenamefont {Raghu}\ \emph {et~al.}(2008)\citenamefont {Raghu},
  \citenamefont {Qi}, \citenamefont {Liu}, \citenamefont {Scalapino},\ and\
  \citenamefont {Zhang}}]{Rag08}%
  \BibitemOpen
  \bibfield  {author} {\bibinfo {author} {\bibfnamefont {S.}~\bibnamefont
  {Raghu}}, \bibinfo {author} {\bibfnamefont {X.-L.}\ \bibnamefont {Qi}},
  \bibinfo {author} {\bibfnamefont {C.-X.}\ \bibnamefont {Liu}}, \bibinfo
  {author} {\bibfnamefont {D.~J.}\ \bibnamefont {Scalapino}}, \ and\ \bibinfo
  {author} {\bibfnamefont {S.-C.}\ \bibnamefont {Zhang}},\ }\href {\doibase
  10.1103/PhysRevB.77.220503} {\bibfield  {journal} {\bibinfo  {journal} {Phys.
  Rev. B}\ }\textbf {\bibinfo {volume} {77}},\ \bibinfo {pages} {220503}
  (\bibinfo {year} {2008})}\BibitemShut {NoStop}%
\bibitem [{\citenamefont {Parish}\ \emph {et~al.}(2008)\citenamefont {Parish},
  \citenamefont {Hu},\ and\ \citenamefont {Bernevig}}]{Par08}%
  \BibitemOpen
  \bibfield  {author} {\bibinfo {author} {\bibfnamefont {M.~M.}\ \bibnamefont
  {Parish}}, \bibinfo {author} {\bibfnamefont {J.}~\bibnamefont {Hu}}, \ and\
  \bibinfo {author} {\bibfnamefont {B.~A.}\ \bibnamefont {Bernevig}},\ }\href
  {\doibase 10.1103/PhysRevB.78.144514} {\bibfield  {journal} {\bibinfo
  {journal} {Phys. Rev. B}\ }\textbf {\bibinfo {volume} {78}},\ \bibinfo
  {pages} {144514} (\bibinfo {year} {2008})}\BibitemShut {NoStop}%
\bibitem [{\citenamefont {Seo}\ \emph {et~al.}(2008)\citenamefont {Seo},
  \citenamefont {Bernevig},\ and\ \citenamefont {Hu}}]{Seo08}%
  \BibitemOpen
  \bibfield  {author} {\bibinfo {author} {\bibfnamefont {K.}~\bibnamefont
  {Seo}}, \bibinfo {author} {\bibfnamefont {B.~A.}\ \bibnamefont {Bernevig}}, \
  and\ \bibinfo {author} {\bibfnamefont {J.}~\bibnamefont {Hu}},\ }\href
  {\doibase 10.1103/PhysRevLett.101.206404} {\bibfield  {journal} {\bibinfo
  {journal} {Phys. Rev. Lett.}\ }\textbf {\bibinfo {volume} {101}},\ \bibinfo
  {pages} {206404} (\bibinfo {year} {2008})}\BibitemShut {NoStop}%
\bibitem [{\citenamefont {Liu}\ \emph {et~al.}(2018)\citenamefont {Liu},
  \citenamefont {Fang}, \citenamefont {Zheng}, \citenamefont {Huang},\ and\
  \citenamefont {Lin}}]{Liu18}%
  \BibitemOpen
  \bibfield  {author} {\bibinfo {author} {\bibfnamefont {G.}~\bibnamefont
  {Liu}}, \bibinfo {author} {\bibfnamefont {S.}~\bibnamefont {Fang}}, \bibinfo
  {author} {\bibfnamefont {X.}~\bibnamefont {Zheng}}, \bibinfo {author}
  {\bibfnamefont {Z.}~\bibnamefont {Huang}}, \ and\ \bibinfo {author}
  {\bibfnamefont {H.}~\bibnamefont {Lin}},\ }\href {\doibase
  10.1088/1361-648x/aae289} {\bibfield  {journal} {\bibinfo  {journal} {J.
  Phys.: Condens. Matter}\ }\textbf {\bibinfo {volume} {30}},\ \bibinfo {pages}
  {445604} (\bibinfo {year} {2018})}\BibitemShut {NoStop}%
\bibitem [{\citenamefont {Yamase}\ and\ \citenamefont {Zeyher}(2013)}]{Yam13}%
  \BibitemOpen
  \bibfield  {author} {\bibinfo {author} {\bibfnamefont {H.}~\bibnamefont
  {Yamase}}\ and\ \bibinfo {author} {\bibfnamefont {R.}~\bibnamefont
  {Zeyher}},\ }\href {\doibase 10.1103/PhysRevB.88.180502} {\bibfield
  {journal} {\bibinfo  {journal} {Phys. Rev. B}\ }\textbf {\bibinfo {volume}
  {88}},\ \bibinfo {pages} {180502} (\bibinfo {year} {2013})}\BibitemShut
  {NoStop}%
\bibitem [{\citenamefont {Dumitrescu}\ \emph {et~al.}(2016)\citenamefont
  {Dumitrescu}, \citenamefont {Serbyn}, \citenamefont {Scalettar},\ and\
  \citenamefont {Vishwanath}}]{Dum16}%
  \BibitemOpen
  \bibfield  {author} {\bibinfo {author} {\bibfnamefont {P.~T.}\ \bibnamefont
  {Dumitrescu}}, \bibinfo {author} {\bibfnamefont {M.}~\bibnamefont {Serbyn}},
  \bibinfo {author} {\bibfnamefont {R.~T.}\ \bibnamefont {Scalettar}}, \ and\
  \bibinfo {author} {\bibfnamefont {A.}~\bibnamefont {Vishwanath}},\ }\href
  {\doibase 10.1103/PhysRevB.94.155127} {\bibfield  {journal} {\bibinfo
  {journal} {Phys. Rev. B}\ }\textbf {\bibinfo {volume} {94}},\ \bibinfo
  {pages} {155127} (\bibinfo {year} {2016})}\BibitemShut {NoStop}%
\bibitem [{\citenamefont {Nica}\ \emph {et~al.}(2017)\citenamefont {Nica},
  \citenamefont {Yu},\ and\ \citenamefont {Si}}]{Nic17}%
  \BibitemOpen
  \bibfield  {author} {\bibinfo {author} {\bibfnamefont {E.~M.}\ \bibnamefont
  {Nica}}, \bibinfo {author} {\bibfnamefont {R.}~\bibnamefont {Yu}}, \ and\
  \bibinfo {author} {\bibfnamefont {Q.}~\bibnamefont {Si}},\ }\href {\doibase
  10.1038/s41535-017-0027-6} {\bibfield  {journal} {\bibinfo  {journal} {npj
  Quantum Mater.}\ }\textbf {\bibinfo {volume} {2}},\ \bibinfo {pages} {24}
  (\bibinfo {year} {2017})}\BibitemShut {NoStop}%
\bibitem [{\citenamefont {Ptok}\ \emph {et~al.}(2020)\citenamefont {Ptok},
  \citenamefont {Kapcia},\ and\ \citenamefont {Piekarz}}]{Pto20}%
  \BibitemOpen
  \bibfield  {author} {\bibinfo {author} {\bibfnamefont {A.}~\bibnamefont
  {Ptok}}, \bibinfo {author} {\bibfnamefont {K.~J.}\ \bibnamefont {Kapcia}}, \
  and\ \bibinfo {author} {\bibfnamefont {P.}~\bibnamefont {Piekarz}},\ }\href
  {\doibase 10.3389/fphy.2020.00284} {\bibfield  {journal} {\bibinfo  {journal}
  {Front. Phys.}\ }\textbf {\bibinfo {volume} {8}} (\bibinfo {year} {2020}),\
  10.3389/fphy.2020.00284}\BibitemShut {NoStop}%
\bibitem [{\citenamefont {Benenti}\ \emph {et~al.}(2017)\citenamefont
  {Benenti}, \citenamefont {Casati}, \citenamefont {Saito},\ and\ \citenamefont
  {Whitney}}]{Ben17}%
  \BibitemOpen
  \bibfield  {author} {\bibinfo {author} {\bibfnamefont {G.}~\bibnamefont
  {Benenti}}, \bibinfo {author} {\bibfnamefont {G.}~\bibnamefont {Casati}},
  \bibinfo {author} {\bibfnamefont {K.}~\bibnamefont {Saito}}, \ and\ \bibinfo
  {author} {\bibfnamefont {R.}~\bibnamefont {Whitney}},\ }\href {\doibase
  https://doi.org/10.1016/j.physrep.2017.05.008} {\bibfield  {journal}
  {\bibinfo  {journal} {Phys. Rep.}\ }\textbf {\bibinfo {volume} {694}},\
  \bibinfo {pages} {1} (\bibinfo {year} {2017})}\BibitemShut {NoStop}%
\end{thebibliography}

%

\end{document}